\documentclass{article}

\usepackage{hyperref}
\usepackage{graphicx}
\usepackage{amsfonts}
\usepackage{url}
\usepackage{caption}
\usepackage{subcaption}
\usepackage{booktabs}
\usepackage{lscape}
\usepackage[table,xcdraw]{xcolor}
\usepackage{amsmath}
\usepackage{wrapfig}

\usepackage{multirow}

\def\acosh{{\rm acosh}}

\begin{document}

\def\bbH{\mathbb{H}}
\def\bbE{\mathbb{E}}
\def\bbS{\mathbb{S}}
\def\bbR{\mathbb{R}}
\def\bbG{\mathbb{G}}
\def\bbA{\mathbb{A}}
\def\bbB{\mathbb{B}}
\def\bbZ{\mathbb{Z}}
\def\bbX{\mathbb{X}}
\def\ra{\rightarrow}

\title{Hyperbolic embeddings for graph compression}
\author{Dorota Celi\'nska-Kopczy\'nska, Eryk Kopczy\'nski \\
Institute of Informatics, University of Warsaw, Warsaw, Poland}

\maketitle

\begin{abstract}
Network theoreticians hypothesize that the structure of real-world networks has a geometric origin.
Especially, hyperbolic geometry was proven insightful in representing and modeling of scale-free networks.
Embedders are algorithms used to find a geometric representation of a network. In this study,
we introduce a fast lossless graph compression algorithm based on modern hyperbolic embedders. Experimental
validation on real-world and generated networks shows that our algorithm beats state-of-the-art by up to 42\% on real-world graphs.
\end{abstract}

\section{Introduction}

Ubiquity of large graphs in today's world prompts a quest to compress such data; that is, for a given graph $G=(V,E)$ where $E \subseteq V^2$, construct a minimal bit sequence $s$ that encodes $E$ losslessly. Beyond minimizing the lenght of sequence $s$, practical requirements include fast compression and decompression and efficient query support (e.g., \emph{what are all the successors of $v \in V$}) with a low working memory footprint \cite{compsurvey}. Designing such representations is central to graph algorithms and succinct data structures.

The dominant practical approach to lossless graph compression is WebGraph \cite{webgraph,webgraph_rs} that relies on the observation that successors of $v$ are often similar to each other, so they are likely to have similar indices (in some natural enumeration of $V$). Instead of writing every successor of $v$ as a full index, we only encode gaps between them using a variable-length encoding. However, WebGraph's effectiveness depends on the ordering of $V$. Consequently, it is typically preceded by computationally expensive reordering algorithms, such as Layered Label Propagation (LLP) \cite{webgraph_llp} or recursive graph bisection (BP) \cite{rgb_bp}.

WebGraph exploits locality from vertex ordering, but does not take into account the latent geometric structure that has proved useful for describing many real-world networks. According to geometric network-generation models, two nodes $v$ and $w$ are likely to be connected if they are in a close neighborhood in a given metric space. 
However, the knowledge of the locations of the nodes in the given (often latent) metric space is needed. A \emph{geometric embedding} of a graph $G$ to a metric space $\bbX$ is a mapping $m: V \ra \bbX$. This way we can operationalize the probability of the connection between two nodes $v$ and $w$ as a function of the distance between $m(v)$ and $m(w)$. The embedding needs to be \emph{good}, typically, we are interested in embeddings maximizing \emph{likelihood}, which measures how good the embedding is at predicting edges between nodes.
An \emph{embedder} is an algorithm which finds a good embedding $m: V \ra \bbX$.

Hyperbolic geometry was found to be a promising choice in visualization and modeling of real-world graphs characterized by similar properties, such as power-law scaling behavior \cite{papa}. In Random Hyperbolic Graph model (RHG) \cite{Krioukov_2010}, the metric space $\bbX$ is a disk of radius $R$ in the hyperbolic plane $\bbH^2$. Every node gets two polar coordinates, $r$ (radial) and $\phi$ (angular). The angular coordinate $\phi$ corresponds to similarity (nodes with close $\phi$ are considered similar) and the radial coordinate $r$ proxies popularity (nodes with small $r$ are considered popular and thus connect to more other nodes, even if they are less similar). For more than a decade, hyperbolic embedders (finding good embeddings $m: V \ra \bbH^2$) have been a vivid research area not only in network theory but algorithmic and machine learning communities as well \cite{bridging_iclr}.

In this paper, we propose a new approach to lossless compression of graphs, based on hyperbolic embeddings. Our contributions are as follows:

\begin{itemize}
\item First practical lossless graph compression algorithm based on hyperbolic embeddings, running in time $O(n B_bB_d+m\log n)$, where $B_b$ and $B_d$ are compression parameters.
 We use \emph{entropy coding} to encode connection in roughly as many bits as the negative binary logarithm of the likelihood. This, together with a succinct
  description of the mapping $m$ itself, yields a graph compression method.
\item Embedding-based vertex ordering can be used as a graph reordering method. CLOVE is especially promising due to its speed and the $\phi$ sequence naturally ordering vertices. Our experiments show that this reordering improves WebGraph's performance as well.
\item Experimental verification using fast embedders: BFKL \cite{tobias}, LPCS \cite{WANG2016609}, HMCS \cite{hmcs}, and CLOVE \cite{hypclove}. Our method outperforms WebGraph-based appoaches by up to 42\% on real-world graphs. The choice of the embedder significantly impacts compression quality.
\end{itemize}

\begin{figure}
\begin{center}
\includegraphics[width=.4\columnwidth]{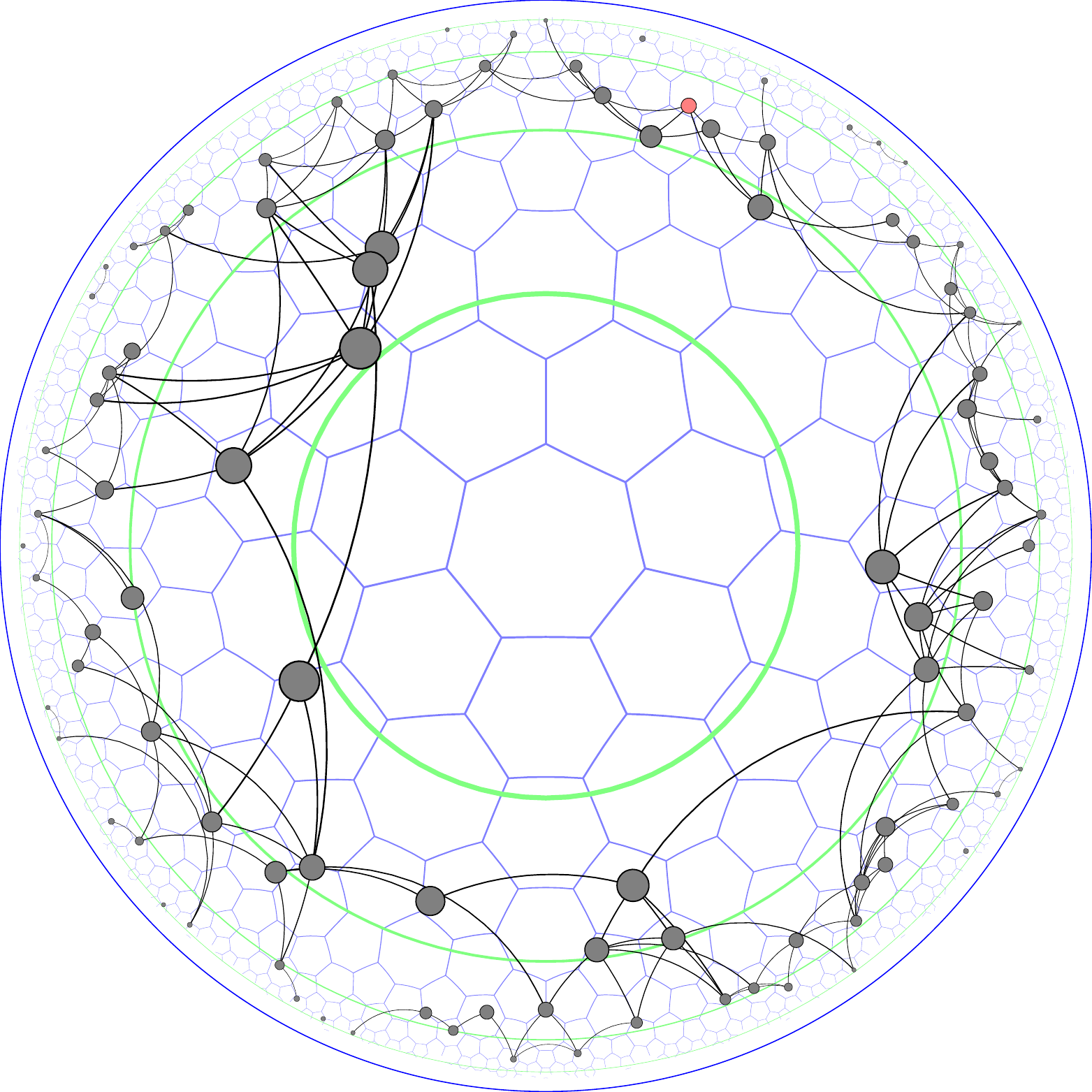}%
\end{center}
\caption{An illustration of a random hyperbolic graph, in the Poincar\'e disk model. All blue edges are of the same hyperbolic length. Edges connected with hyperbolic straight lines.
Green circles have radii 1, 2, 3, 4. \label{fig:coviz1}}
\end{figure}

\section{Prerequisites}\label{sec:prereq}
In this paper, we work with embeddings of graphs into the hyperbolic plane $\bbH^2$. For more information on this geometry, see, e.g., the book \cite{cannon}; we also recommend game HyperRogue \cite{hyperrogue} to gain intuitions.

Let $G = (V, E)$, where $E \subseteq V^2$, be a graph. We denote $n=|V|$, $m=|E|$.
Each $v \in V$ is mapped to a point $m(v) \in \bbH^2$ with polar coordinates $r(v)$ and $\phi(v)$. 
Figure \ref{fig:coviz1} illustrates an example embedded graph.
The hyperbolic distance between two points $(r_1,\phi_1)$ and $(r_2,\phi_2)$,
where $\phi_1-\phi_2=\phi$ is given by the following formula:

\begin{eqnarray}
d &=& \acosh\ (\sinh(r_1)\sinh(r_2)\cos\phi + \cosh(r_1)\cosh(r_2)) \\
  &=& \acosh\ (\cosh(r_1-r_2) + (1-\cos(\phi))\sinh(r_1)\sinh(r_2)) \label{precisedist}
\end{eqnarray}

According to the RHG model \cite{Krioukov_2010}, the expected probability of two nodes $(v,w)$ being connected is $p(v,w)=p(\delta(v,w)) = (1 + \exp((\delta(v, w)-R))/2T))^{-1}$,
where $R$ is the radius of the disk, and $T$ (called \emph{temperature}) is a parameter of the model controlling how well the latent geometry reflects connections.
We measure the quality of a hyperbolic embedding with \emph{likelihood} $L$, which is the product of $p(v,w)$ for all pairs of nodes $(v,w) \in E$, and $1-p(v,w)$ for
all pairs of nodes $(v,w) \notin E$. While the parameters $R$ and $T$ have specific meanings in the RHG model, for compression purposes, we simply pick the values of $R$ and $T$
which yield the maximum likelihood.

\subsection{Entropy coding}\label{sec:entcod}
We can use a compression method similar to arithmetic encoding \cite{ari_enc} to create a compression method based on a good quality hyperbolic embedding $m$. We start with the interval
$[a,b] = [0,1]$. To compress an information about an event $A$ that happens with probability $p$, we compute $m = a + (b-a)p$, and then replace the interval with $[a,b]$ with $[a,m]$
if $A$ happened, or with $[m,b]$ if not. We repeat this procedure for every event we are interested in; to compress the edge information in a hyperbolic embedding, we simply repeat it
for every pair $(u,v)$. The compressed string of bits is the binary representation of some simplest number in the final interval $[a,b]$. It is easy to see that the length of the final
interval equals $L$, and therefore, the length of the compressed string is $O(1)-\log_2 L$.
Thus, we can obtain a short representation of the connection structure.

We also need to represent the embedding itself. How to do so best may depend on the embedding; in general, distances
in hyperbolic geometry are extremely sensitive to small differences in coordinates \cite{dhrg_sea,achilles,numeric_pub}, so we may require 
a large number of bits to represent their coordinates.

\subsection{Quick access to successor lists}\label{sec:randacc}
While the entropy coding method described above gives a compression as good as the information-theoretic limit, we also need to care about practical issues. We want the compression
and decompression to be quick (e.g., linear in the length of the output). In graph compression, we also usually not only want to be able to recover the
information about the whole graph, but also to get a list of successors of a given node $v$.
Perfect arithmetic coding is impractical due to using arbitrary precision real numbers, on which computations are slow.

These issues can be solved by using an implementation in which we represent only next $Q$ bits which are not yet decided (and thus not yet sent to the
output stream), i.e., these bits differ in the numbers $a$ and $b$. We can use a small value of $Q$ (e.g., $Q=20$) and obtain very fast compression close to perfect.
Now, for every node $u$, we encode every potential edge $u \rightarrow v$ in sequence. Before we do so, we record the current offset in the compressed stream, as well as
the current bits of $a$ and $b$. This information lets us answer the successor list queries by moving to the recorded position in the compressed stream and interval,
and reading the edges for all $u$. We can also use separate runs of entropy coding to encode every $u$ -- this yields shorter offset lists and a possibility of
parallel computation, at the cost of $n$ extra bits.

\subsection{Undirected, directed, and bipartite graphs}\label{sec:directed}
In research on graph compression graphs are assumed to be directed, contrary to most works on hyperbolic embeddings, on undirected graphs.
Ongoing research aims to fill this gap. In \cite{Jankowski_bipartite}, for bipartite graphs, assume $V = V_1 \cup V_2$, and all edges go from $V_1$ to $V_2$.
Since we only consider pair of nodes $(v,w)$ in which $v \in V_1$ and $w \in V_2$ as possible edges, we take only them into account in the formula for likelihood.
In \cite{diremb} directed graphs are embedded by splitting every node that has both in-edges and out-edges into two (in-node and out-node).

However, for compression purposes we need fast embedders, and those known to us assume undirected graphs. In our method we run a fast hyperbolic
embedder on the undirected graph $(V,E')$ such that $\{u,v\}\in E'$ if and only if $(u,v) \in E$ or $(v,u) \in E$.
To correctly take directed or bipartite graphs into account, we then introduce changes based on the ideas from \cite{Jankowski_bipartite,diremb}.

Many graphs we benchmark our algorithm on are in fact undirected ($(u,v) \in E$ iff $(v,u) \in E$), or have high reciprocity. Theoretically
we could use this information to greatly reduce the length of the compressed string. We do not do so, because this is not usually done in WebGraph-based graph compression algorithms,
and also doing so prevents answering successor list queries efficiently.

\begin{figure}
\begin{center}
\includegraphics[width=.4\columnwidth]{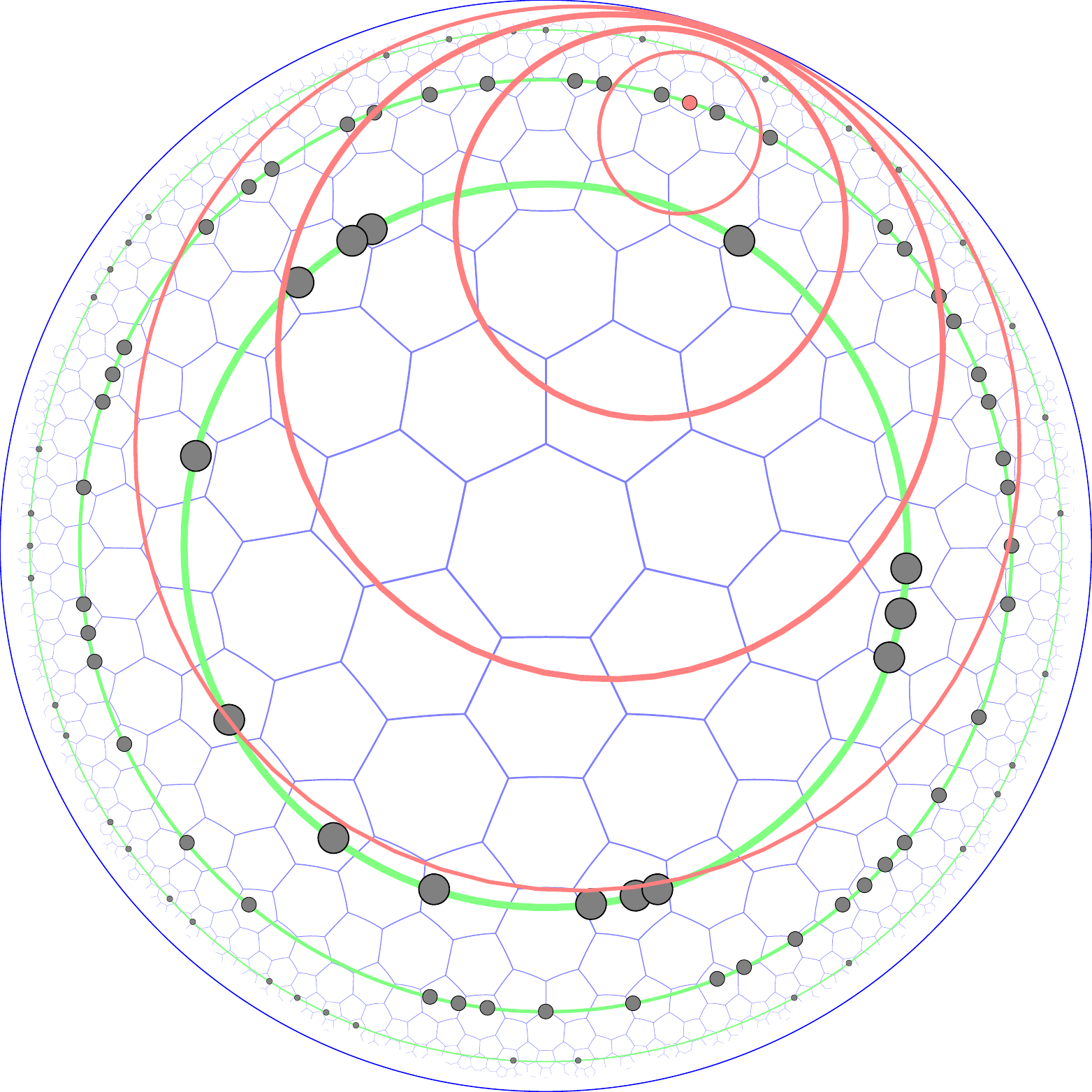}%
\end{center}
\caption{An illustration of the HyperFast algorithm. The radial buckets are $[1,2)$, $[2,3)$ and $[3,4)$. We change the radial coordinate of every nodes from Figure \ref{fig:coviz1} 
to the average radial distance in its bucket. Distance buckets are $[0,1)$, $[1, 2)$, $[2,3)$, $[3, 4)$ and $[4, \infty)$. Red circles show which nodes fall into which distance bucket with respect
to the red node. \label{fig:coviz2}}
\end{figure}

\section{Fast compression algorithm}\label{sec:hyperfast}

The method described so far is slow. We need to separately encode each of $n^2$ edges, and thus, the compression algorithm runs in time $O(n^2)$
and can obtain the list of successors of $v \in V$ in time $O(n)$.
In this section, we describe a more efficient method, called HyperFast\footnote{Placeholder name, for anonymization purposes.}. Instead of computing the
exact distances for every pair of nodes, we approximate without significantly worsening the likelihood.

We assume that the nodes are already ordered by the angular coordinate $\phi$. To simplify the compression, we assume that $i$-th node has $\phi_i = 2\pi i/n$ -- this is true in
a CLOVE embedding \cite{hypclove} and hopefully will not worsen the likelihood too much otherwise.

The method is based on approximations. We split the set of radial distances $A_r = \{r_v: v \in V\}$ into $B_r$ intervals, that we will call radial buckets. In our basic implementation,
we split the interval between the $\min(A_r)$ and $\max(A_r)$ into $B_r$ equal intervals; let $\bbB_r$ be the set of all radial buckets. For every $v \in V$, let $b_v$ be the
radial bucket that $r_v$ belongs to. For every bucket $b \in \bbB_r$, let $S_b$ be the set of nodes that fall into radial bucket $b$ (geometrically, an annulus), and $\hbar{r_b}$
be the average of all $r_v$ for all $v \in S_b$. We replace every $r_v$ with $\hbar{r_{b_v}}$, thus compressing the annulus $S_b$ into a circle (Figure \ref{fig:coviz2});
thus, we lose some precision of radial distances, but hopefully that should not matter much for the quality of edge prediction.
We send the following data to the compressed stream: the $\hbar{r_b}$ for every radial bucket
$b$, and $b_v$ for every node $v \in V$. We use arithmetic encoding for sending $b_v$, since the distribution of nodes into radial buckets is not uniform (in typical hyperbolic
embeddings significantly more nodes fall into buckets with greater indices).
This way, we have compressed all the information about radial distances into $B_r$ floating-point numbers and $O(n \log(B_r))$ bits.
In Figure \ref{fig:coviz2}, the radial buckets are $[1,2)$, $[2,3)$ and $[3,4)$.

Now, let $A_d = \{d(u,v): (u,v) \in E\}$ be the set of distances between vertices connected with edges. We split the set of possible distances into $B_d$ intervals,
that we will call distance buckets. In our basic implementation, we again split the interval between $\min(A_d)$ and $\max(A_d)$ into $B_d-1$ equal buckets,
and another bucket for distances above $\max(A_d)$ (that can be achieved for nodes not connected with edges). Distances below $\min(A_d)$ can be considered a part of the first
instance bucket. Let $\bbB_d$ be the set of all distance buckets. The idea of this bucketing is that we will no longer care about the exact distance $d(u,v)$, but what bucket
$c \in \bbB_d$ it belongs to. The compressor sends the information about the buckets obtained ($B_d$ threshold distances, which are floating point numbers).
In Figure \ref{fig:coviz2}, distance buckets are $[0,1)$, $[1, 2)$, $[2,3)$, $[3, 4)$ and $[4, \infty)$. Red circles show which nodes fall into which distance bucket with respect
to the red node.

For every $u \in V$, every radial bucket $b \in \bbB$, and every distance bucket $c$, the set of nodes $v$ which are in bucket $b$ and such that $d(u,v) \in c$ has a very simple structure.
It is an intersection of the circle $S_b$ and the annulus $X_{u,c}$ of all points $v$ whose distance from $u$ is in the distance bucket $c$ (Figure \ref{fig:coviz2}).
Since $V$ was ordered by the angular coordinate, so is the circle $S_b$, and thus we can find the two arcs of $S_b$ which intersect the annulus $X_{u,c}$ using binary search.
It is also possible to remove the logarithmic factor by using the two-pointer caterpillar method. To do so, we need to compute 
$S_b \cap X_{u,c}$ for all the source vertices $u$ in the given radial bucket in their angular order.

By performing the construction above for every $u \in E$, radial bucket $b \in \bbB_r$ and distance bucket $c \in \bbB_d$, we can compute $m'_c$, the number of pairs $(u,v) \in V \times V$
such that $d(u,v) \in c$. This computation can be done using only the information we have sent/received so far, in time $O(nB_bB_d \log n)$, or $O(nB_bB_d)$ using the two-pointer
caterpillar method.

Knowing the graph, we can also, for every distance bucket $c \in \bbB_d$, compute $m_c$, the number of actual edges $(u,v) \in E$ such that $d(u,v) \in c$.
We send this information to the compressed stream. Since $0 \leq m_c \leq m'_c \leq n^2$, this requires $O(B_d \log n)$ bits.

Now, for every distance bucket $c$, we need to send the information about specifically which of these $m'_c$ elements correspond to edges.
Denote $p_c = \frac{m_c}{m'_c}$. For simplicity, we perform the compression here assuming that, for each $c$ and for each of $m'_c$ edge pairs, they are connected with an edge independently
with probability $p_c$. We can thus encode the information about the connections using $\sum_{c \in \bbB_d} m_c H(p_c)$ bits, where $H(p)$ is the entropy $-p \log(p) +-(1-p) \log(1-p)$.
If the radial and distance buckets are small enough, this value should be close to $-\log L$ (in fact, it could be smaller, since the actual probability $p(d)$ need not actually be a logistic function of $d$).

We again do this separately for every $u \in V$, radial bucket $b \in \bbB$, and distance bucket $c$. Recall that $S_b \cap X_{u,c}$ is a union of two arcs; we also handle these two
arcs separately. Let $S_b \cap X_{u,c} = \{v_1, \ldots, v_j\}$, and let $I$ be the set of indices which correspond to $v \in S_b$ such that $(u,v) \in E$. 
Observe that  when $p$ is very small (as it is for all buckets which contain mostly non-edges), there are very long gaps between the elements of $I$.
We send the information about the following random event: \emph{is I empty?} Since we know the probability of this event according to
our independence model, $(1-p)^j$, we can use the entropy coding method detailed in Subsection \ref{sec:entcod} to encode this information nearly optimally. In case if
the event happened (the intersection was empty), there is nothing to do anymore. Otherwise, we use binary search to pinpoint the specific index $\min(I)$.
The invariant of binary search is $\min(I) \in \{l, \ldots, r\}$, initially $l=1$ and $r=j$; we pick $s = (l+r)/2$, and consider the event $\min(I) \leq s$.
We can compute the probability of this event using our independence assumption, and thus, we can use the entropy conding method to encode this information.
Depending on whether the event holds or not, we obtain smaller and smaller intervals, until we pinpoint $\min(I)$.
Taking $K=n$, the compression/decompression algorithm, for a given u, needs $O(\deg u \log(n) + B_d B_r)$ queries to entropy coding, and has the same time complexity.
Here, $\deg u$ is the outdegree of $u$.

\paragraph{Adjustments.}
The basic HyperFast algorithm outlined above can be modified, e.g., by optimizing radial and distance buckets, using directed degree-based radial distances, triangle-based
edge prediction, and local search. These approaches are described in detail in Section \ref{sec:furthopt}.

\section{Experimental setup}\label{sec:experimental}
To evaluate our idea for practical uses, 
we analyze the results of the compression methods on real-world and simulated graphs.
For each graph $G=(V,E)$ we compute the bits per link for:

\begin{itemize}
\item the \emph{trivial} representation of $G$: each one of $n(n-1)$ potential edges is independently encoded
using arithmetic encoding, under assumption that the probability of an edge occuring is $\frac{m}{n(n-1)}$. This is a natural choice for the baseline.
\item the WebGraph representation of $G$, ordering by the nodes' order of appearance in the input file (``input order'');
\item the WebGraph representation of $G$, ordering by the recursive graph bisection method (BP) \cite{rgb_bp}.
We use two variants of selecting the original bisection: random (BP) or based on the input order (BPN);
\item the WebGraph representation of $G$, ordering by the layered label propagation (LLP) \cite{webgraph_llp};
\item the WebGraph representation of $G$, ordering by the $\phi$ order from a given hyperbolic embedder;
\item the HyperFast representation of $G$, based on a hyperbolic embedder, for two sets of hyperparameters: $B_d=B_r=8$ and $B_d=B_r=16$;
\end{itemize}

We use the following hyperbolic embedders: CLOVE \cite{hypclove}, BFKL \cite{tobias}, HMCS \cite{hmcs}, and LPCS \cite{WANG2016609}.

The complete list of real-world graphs we use can be found in Section \ref{app:networks}.
Our set of real-world graphs contains popular benchmarks used in the studies on both hyperbolic embedders and the state-of-the-art graph compression methods. To generalize our findings, we also work with the simulated graphs from the following generative graph models. In each case, we generate graphs of approximate size $n=10000$ and $n=50000$, and an approximate average degree of 10.

\begin{itemize}
\item {\bf er}: Erd\"os-Renyi model \cite{erdosrenyi}. Every pair of vertices is connected with the same probability.
\item {\bf ab}: Albert-Barabasi model \cite{albertbarabasi}. Graphs with power law degree distribution ($\gamma=3$).
\item {\bf cl}: Chung-Lu model \cite{chunglu,chunglu_powerlaw}. Graphs with the given expected degree distribution. We use the power law degree distribution ($\gamma=2.5$).
\item {\bf gs}: a simple geometric model. Nodes are randomly set on a sphere and connected if the distance is shorter than the threshold.
\item {\bf rh}: RHG model. We use the implementation from \cite{tobias}. This implementation does not allow setting the average degree directly, but we still keep average degree close to 10.
\item {\bf np}: nPSO model \cite{npso}. This model is similar to RHG, but the angular coordinate is not uniform, to allow creating graphs with requested number and relative size of communities.
We create graphs with 8 communities, with their relative sizes chosen randomly.
\end{itemize}

According to Network Theory, many real-world networks possess similar characteristics, especially the power-law degree distributions (with $\gamma$ values within (2,3)). A common hypothesis is that the nodes connect based on their similarity. If we assume that such a similarity can be modeled with geometry, i.e., the distance between the nodes proxies the probability of their connection, we get the foundation assumption of the \emph{geometric} generative graph models; in our setup those are gs, rh, and np models. Those models are shown to produce networks that resemble real-world ones both in the degree distribution and connectivity. On the contrary, ab and cl models are \emph{non-geometric} alternatives to produce networks with power-law degree distribution. Eventually, er model is a popular benchmark in Network Theory in which nodes do not connect based on their similarity. For the reproduciblity purposes, in the case of rh models, we follow the setups from comparative studies on hyperbolic embedders \cite{bridging_iclr}.
RHG model creates ultra-small world networks \cite{bogunya_division} (the average distance grows slower than any polynomial of the logarithm of the size of the network: heterogeneous networks with presence of the hubs).




Our preliminary analysis (Section \ref{sec:furthopt}) shows that the following adjustments to the basic HyperFast algorithm yield good results on real-world graphs, in
terms of the the compression-to-time tradeoff: optimizing radial and
distance buckets, using directed degree-based radial distances, and triangle-based edge prediction. In the next two sections these adjustments are applied.

\section{Analysis of real-world graphs}\label{sec:realworld}

\paragraph{Randomness in embedders}

Most embedders are randomized, so we repeated a portion of the experiments with different seeds. For each real-world network, we ran 10 iterations per randomized algorithm (excluding slower embedders on the largest graphs).
Our result suggest that randomness does not threaten the validity of our approach. Coefficients of variations across all embedder-based approaches are very low (in most cases they do not exceed one percent), suggesting high homogeneity of the results (see Section \ref{app:tables}).

\paragraph{Overall quality assessment}

\begin{figure}
  \begin{center}\includegraphics[width=0.4\columnwidth]{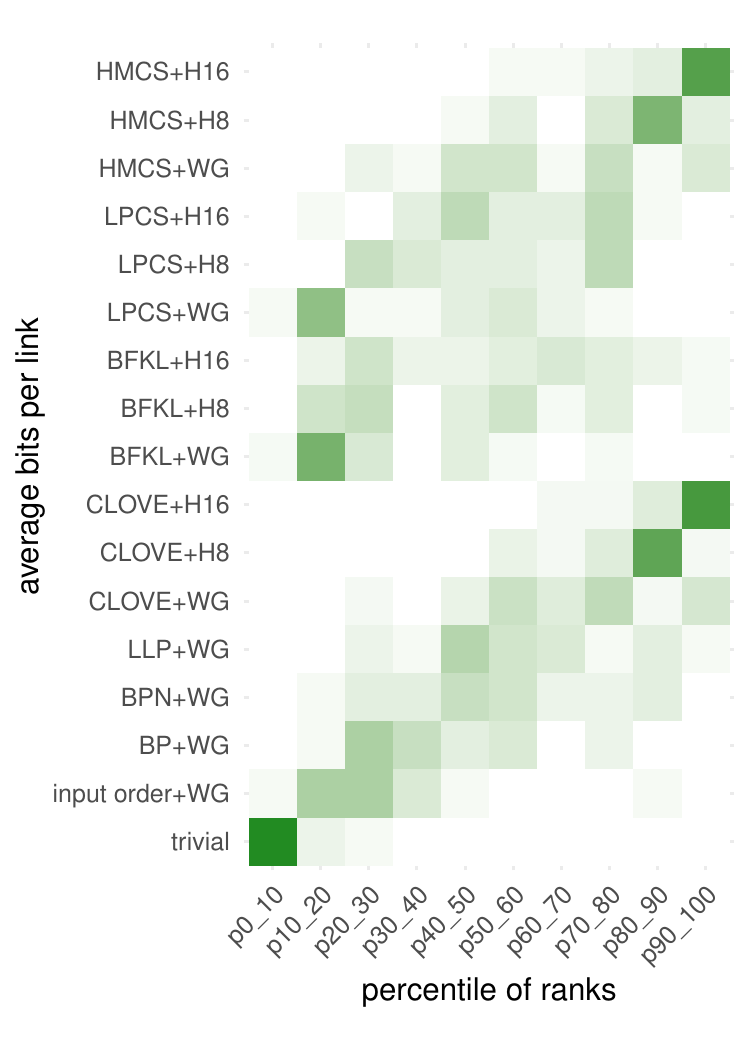}
  \includegraphics[width=0.4\columnwidth]{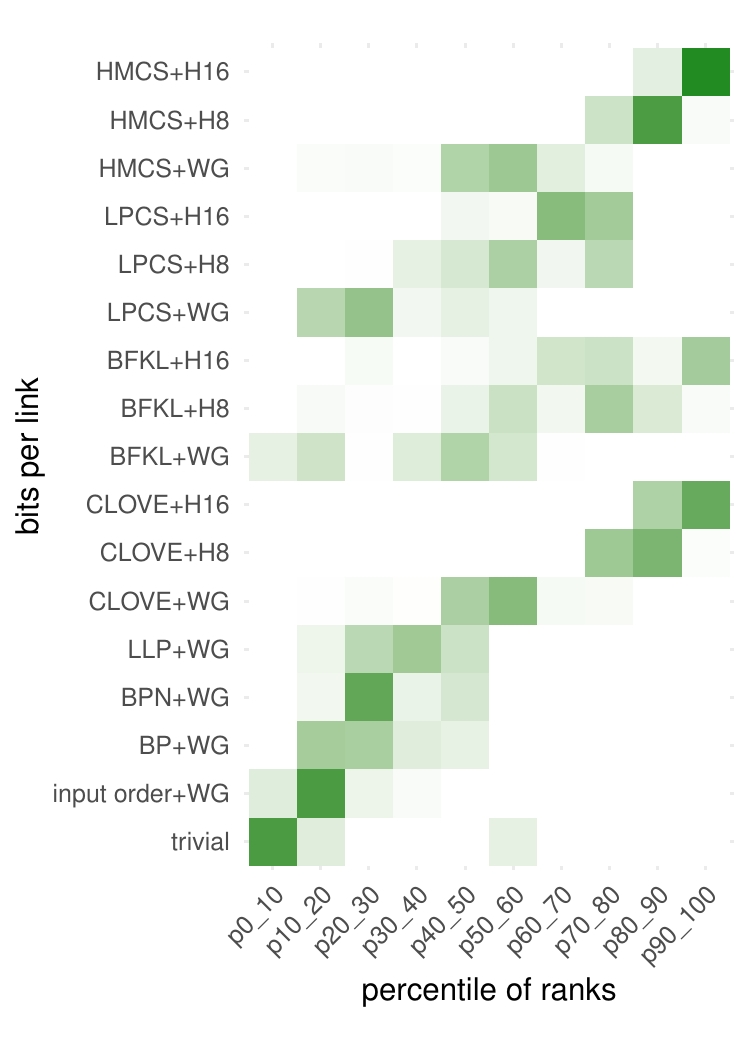}\end{center}
  \caption{Quality assessment of compression methods on real-world (left) and simulated (right) graphs. 
Darker colors indicate that the given method occurred more frequently in the
given percentile of ranks (higher percentiles are better) over all graphs
benchmarked.
\label{fig:realworld}\label{fig:simulated}}
\end{figure}

We ranked methods by their average BPL across 10 iterations for each graph, with lower BPL yielding higher ranks. Figure \ref{fig:realworld} depicts the aggregate results, while Section \ref{app:tables} contains plots and tables.

Our analysis suggests that HyperFast with HMCS or CLOVE (especially with $B_d=B_r=16$) generally outperforms non-embedder approaches, achieving higher (better) ranks more frequently (Fig. \ref{fig:realworld}). Surprisingly, methods utilizing BFKL or LPCS usually do not perform well. This is slightly unfortunate, as those embedders are among the fastest ones. HMCS appears promising here as it compromises compression quality with the computation time. The current state-of-the-art compression method (WebGraph with LLP) turns out to be a mediocre choice when it comes to the quality of the compression. It usually underperforms CLOVE or HMCS approaches, but it is usually better than the approaches based on BFKL or LPCS. 

\paragraph{Maximum performance}

Mean BPL offers a baseline expection for each method. However, one may be also interested in the best (optimistic) scenarios. To this end, we provide additional analyses on the best (minimal) BPL achieved per graph (where applicable). The rankings remain practically unchanged whether based on means or the best-case values (weighted Cohen's kappa 0.99, indicating nearly perfect agreement). 

\begin{table*}
\begin{center}
\begin{tabular}{|l|l|r|r|r|r|r|r|r|r|}
\hline
embe- & algo & \multicolumn{4}{c|}{vs trivial} & \multicolumn{4}{c|}{vs LLP} \\ \cline{3-10}
dder  & rithm & min & avg & max & sd & min & avg & max & sd \\ \hline
BFKL  & WG    & -0.84 & 34.14 & 100   & 23.36 & -53.86 & -19.53 & 100 & 31.32 \\
CLOVE & WG    & 12.75 & 47.53 & 78.75 & 17.5 & -7.83 & 4.52 & 21.55 & 7.5 \\
HMCS  & WG    &  2.49 & 46.56 & 92.25 & 20.88 & -8.68 & 3.32 & 19.63 & 6.94 \\
LPCS  & WG    & -2.12 & 39.01 & 87.29 & 22.44 & -53.12 & -12.65 & 15.5 & 15.58 \\ \hline
BFKL  & H8    & 16.59 & 42.55 & 62.56 & 13.08 & -351.06 & -24.13 & 24.65 & 75.61 \\
CLOVE & H8    & 33.84 & 55.05 & 75.04 & 11.48 & -13.68 & 15.37 & 38.18 & 14.74 \\
HMCS  & H8    & 26.61 & 54.25 & 82.71 & 13.6 & -108.32 & 10.12 & 36.79 & 28.25 \\
LPCS  & H8    & 23.82 & 47.39 & 77.19 & 14.5 & -174.83 & -5.34 & 34.85 & 39.97 \\ \hline
BFKL  & H16   & 18.93 & 44.38 & 66.22 & 13.48 & -307.01 & -18.62 & 26.57 & 67.21 \\
CLOVE & H16   & 33.85 & 57.08 & 77.72 & 12.07 & -8.06 & 19.78 & 42.17 & 13.4 \\
HMCS  & H16   & 28.06 & 56.26 & 86.25 & 14.25 & -65.66 & 15.81 & 41.17 & 21.1 \\  
LPCS  & H16   & 24.52 & 49.17 & 80.68 & 15.22 & -132.82 & -0.05 & 37.79 & 32.57 \\ \hline 
\end{tabular}
\end{center}
\caption{Statistics on maximum performance}\label{stat_max_perf}
\end{table*}

Based on Table \ref{stat_max_perf}, we notice that the embedder-based methods yield substantial improvements over trivial compression. In optimistic scenarios, improvements range from about 60\% to as much as 80\% in the extreme cases, though WebGraph with BFKL or LPCS occasionally shows marginal declines. Against current SOTA (LLP), CLOVE or HMCS in HyperFast with $B_d=B_r=16$ achieves up to about 42\% improvement in optimistic scenarios, as in {\sc ias} graph.
However, 
we observed a small decline compared to LLP+WebGraph on {\sc hepph} (5\%) and {\sc astroph} (8\%), and a significant decline on {\sc uk-2002} (66\%).

\paragraph{Time comparison}

\begin{figure*}
\begin{center}
\begin{tabular}{|l|r|r||r|r|r|r|r|r|} \hline 
graph name&$n$&$m$&WG&LLP&HMCS&seg&comp&decomp \\  \hline 
uk-2014-tpd          &    1766010 &   16861750 & 3.2s & 40.5s & 294.5s & 5.7s & 14.6s & 12.1s \\  \hline
fork-10              &    2480676 &   47860777 & 12.0s & 266.5s & 939.7s & 14.5s & 51.5s & 41.2s \\  \hline
patents              &    3764117 &   16511740 & 3.0s & 63.8s & 917.4s & 21.6s & 18.3s & 16.1s \\  \hline
google               &     855802 &    5066842 & 0.9s & 9.0s & 202.5s & 2.1s & 4.2s & 3.7s \\  \hline
\end{tabular}

\end{center}
\caption{Time breakdown on big real-world graphs.\label{timecomp}}
\end{figure*}

Table \ref{timecomp} shows the runtime breakdown for each method. The time to compress the graph using LLP is the sum of LLP+WG columns; for HMCS+WG,
the sum of HMCS+WG columns; for HMCS+HyperFast with 16 buckets, the sum of HMCS, segmentation, and compression time columns. For decompression, LLP and HMCS columns do not matter,
and the compression time is replaced by the decompression time.

\section{Analysis of the artificial networks}\label{sec:artworld}

For a robust statistical inference on our method's properties, we performed simulation-based evaluation using 2,800 geometric, non-geometric, and random networks. 

\paragraph{Ranks}
The ranks' distribution in Figure ref{fig:simulated} is similar to the one reported for the real-world graphs. Consistent with our findings on real-world networks, our method performs better on larger networks.
HyperFast with HMCS outperforms competing methods across all hyperparameters' configurations. HyperFast with CLOVE performs slightly worse than the HMCS variants. An interesting case is the BFKL embedder: it achieves reasonable gains relative to other methods on most graphs but underperforms significantly on spherical geometry-generated graphs. The current SOTA (LLP+WebGraph) is again a mediocre choice in terms of BPL-based compression quality. 

\paragraph{Does usage of embedders yield significant improvement?}
To adress this question we performed paired Wilcoxon tests with the Bonferroni correction for the number of comparisons. See Section \ref{app:inference} for the full results.
Our results suggest that the differences between the results obtained from HyperFast and the WebGraph based methods are significantly higher than zero on most graph types, especially those resembling real-world networks. The only exceptions occured with LPCS ($B_d=B_r=8$) when compared against WebGraph with CLOVE or HMCS (HyperFast performed significantly worse here). This result is not worrying though given prior observations that LPCS may underperform relative to HMCS or CLOVE on real-world networks. Notably, our proposition beats SOTA (LLP+WebGraph) in every scenario.

Comparisons among WebGraph-based methods are less clear-cut. CLOVE usually significantly improves the results. HMCS does not significantly improve performance on non-geometric graphs but consistently outperforms LLP+WebGraph. AS with real-world networks analysis, BFKL or LPCS usage may not be reasonable. However, we want to stress, that in the ultra-small-world regime (that covers most of the real-world graphs), BFKL yields significantly better results than the WebGraph-based methods (only insignificant difference with WebGraph+LLP).

\paragraph{HyperFast vs WebGraph}

\begin{figure}[h]
  \begin{center}\includegraphics[width=.6\columnwidth]{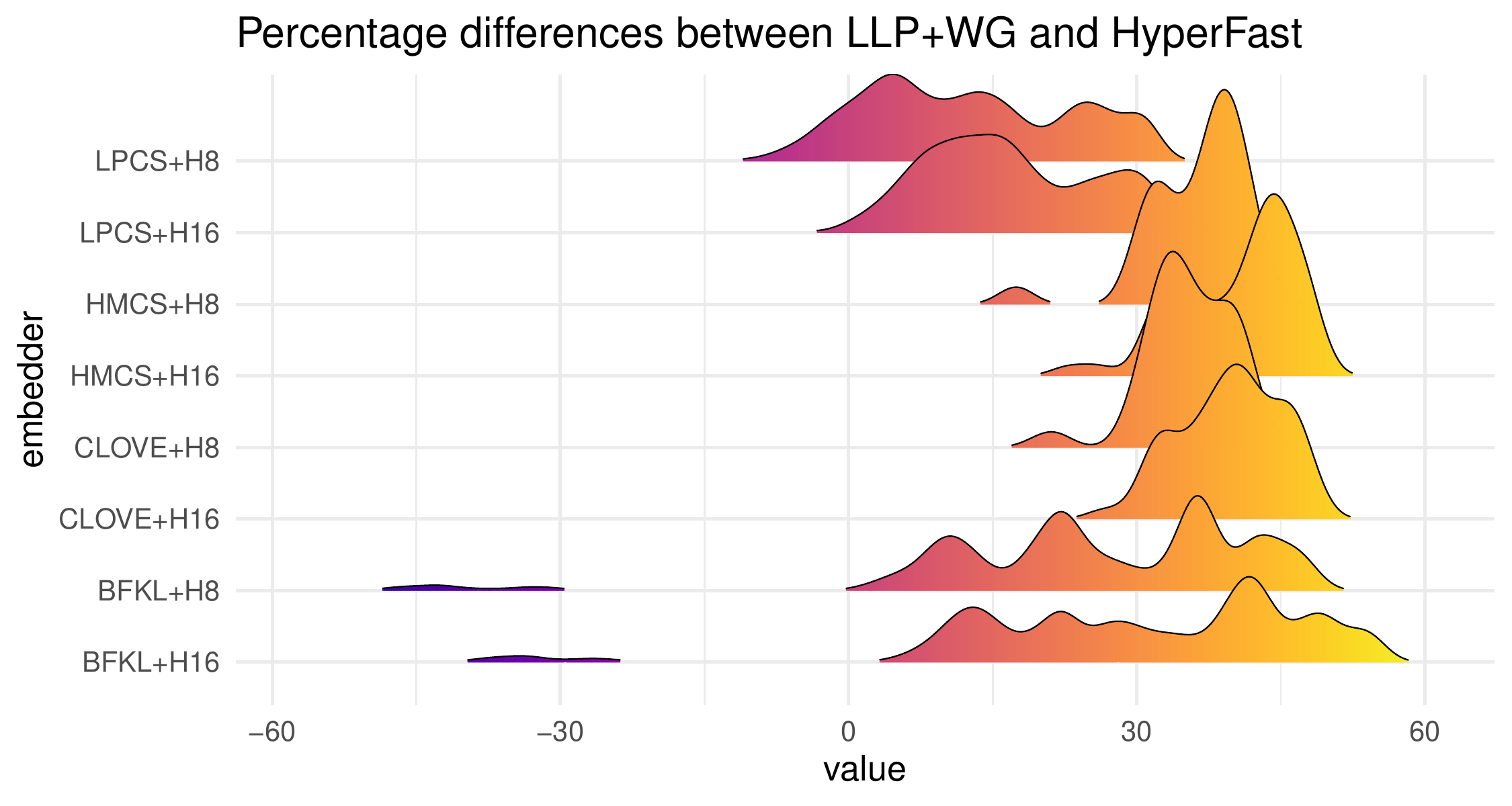}\end{center}
  \caption{Density of the percent difference between the BPL for HyperFast against the SOTA (LLP+WebGraph). Positive values of the percentages indicate the improvement in BPL due to usage of HyperFast. \label{densities}}
\end{figure}

Based on the plots in Figure \ref{densities}, we notice that excluding BFKL, HyperFast allows us to gain improvement over SOTA up to even over 50 percent. Especially, HMCS and CLOVE density curves peak about 40\%. Similarly to the insights from the real-world graph analysis, LPCS or BFKL usage is not recommended.

\paragraph{Factors explaining improvement}
To identify factors driving differences between results from HyperFast against the SOTA, we introduce four regression models setups:

\begin{itemize}
\item {\bf setup:} the basic model controling for the type of the \emph{embedder} (base: BFKL), the number of \emph{nodes}, and \emph{edges} of the graph, whether we used \emph{h16} as the value for the hyperparameters for HyperFast, and the name of the \emph{model} (base: er model).
\item {\bf network-based:} variables computable without the model knowledge. Excludes model name; instead introduces network characterisics: \emph{gcc}, and \emph{acc} controling for global and average clustering coefficient; \emph{md} -- mean distance; \emph{dc} -- average degree; \emph{ecomms} -- the number of communities found by Levene algorithm, \emph{ci} -- average community intimacy, \emph{div} -- KL divergence measuring community size heterogeneity relative to equal sizes;
\item {\bf model-based:} similar to setup, but replaces model name with model-specific variables: \emph{beta} $= 1/(\gamma -1)$, model \emph{temperature} (with interaction for models including it), and \emph{linearity}.
Linearity measures how well a single-dimensional geometry describes the connections in the network; this is 0 for non-geometric models, and otherwise, the inverse of the Hausdorff dimension
of the geometric space (1 for np and rh models, $\frac{1}{2}$ for gs since a single dimension only partially explains the two-dimensional spherical space).
\item {\bf only hyperbolic:} same as network-based, but only on hyperbolic models (rh and np). Includes CScore of the embedding (higher values indicate better angular covariate preservation).
\end{itemize}

Due to severe multicolinearity of variables (e.g., temperature is correlated with clustering coefficient, graph diameter is highly correlated with the mean distance,...) it is not reasonable to put all of them in a single regression. As a part of the variability is related to the factors influencing LLP+WebGraph performance that we do not control for, for the further understanding, we also performed those regressions against trivial compression as the baseline, getting stylized facts about expected performance of our method.

\begin{table*}
\resizebox{\textwidth}{!}{
\begin{tabular}{|l|r|r|r|r|r|r|r|r|r|r|r|r|r|r|r|r|}
\hline
  & \multicolumn{8}{c|}{versus LLP} & \multicolumn{8}{c|}{versus trivial} \\ \hline
  & \multicolumn{2}{c|}{setup} & \multicolumn{2}{c|}{network-based} & \multicolumn{2}{c|}{model-based}   & \multicolumn{2}{c|}{only hyperbolic} 
  & \multicolumn{2}{c|}{setup} & \multicolumn{2}{c|}{network-based} & \multicolumn{2}{c|}{model-based}   & \multicolumn{2}{c|}{only hyperbolic} 
\\ \hline
                               & Est.   &  p      & Est.   & p     & Est.    & p     & Est.     &  p     & Est.   &  p     & Est.   & p       & Est.   & p      & Est.     &  p  \\ \hline
(Intercept)                    &  23.84 &   0.00  &  19.90 &  0.00 &  25.20  &  0.00 &  -10.47  &   0.00 &  3.17  &  0.00  &  29.23 &    0.00 &   7.31 &   0.00 &  43.58   &   0.00 \\
CLOVE                          &   9.95 &   0.00  &   9.95 &  0.00 &   9.95  &  0.00 &    6.85  &   0.00 &  6.90  &  0.00  &   6.90 &    0.00 &   6.90 &   0.00 &   5.17   &   0.00 \\
HMCS                           &  10.83 &   0.00  &  10.83 &  0.00 &  10.83  &  0.00 &    9.12  &   0.00 &  7.41  &  0.00  &   7.41 &    0.00 &   7.41 &   0.00 &   6.37   &   0.00 \\
LPCS                           & -12.80 &   0.00  & -12.80 &  0.00 & -12.80  &  0.00 &  -18.18  &   0.00 & -4.69  &  0.00  &  -4.69 &    0.00 &  -4.69 &   0.00 &  -7.37   & - 0.00 \\
nodes                          &   0.00 &   0.00  &   0.00 &  0.01 &  -0.00  &  0.00 &    0.00  &   0.00 &  0.00  &  0.00  &  -0.00 &    0.00 &  -0.00 &   0.00 &  -0.00   &   0.00 \\
edges                          &  -0.00 &   0.00  &  -0.00 &  0.07 &   0.00  &  0.00 &   -0.00  &   0.06 & -0.00  &  0.00  &   0.00 &    0.00 &   0.00 &   0.00 &   0.00   &   0.00 \\
h16                            &   4.43 &   0.00  &   4.43 &  0.00 &   4.43  &  0.00 &    4.83  &   0.00 &  2.18  &  0.00  &   2.18 &    0.00 &   2.18 &   0.00 &   2.39   &   0.00 \\
model: ab                      &  -0.43 &   0.36  &        &       &         &       &          &        &  4.92  &  0.00  &        &         &        &        &          &        \\
model: cl                      &  -2.44 &   0.00  &        &       &         &       &          &        & 13.08  &  0.00  &        &         &        &        &          &        \\
model: gs                      & -24.75 &   0.00  &        &       &         &       &          &        & 55.29  &  0.00  &        &         &        &        &          &        \\
model: np                      &  -0.66 &   0.06  &        &       &         &       &          &        & 43.73  &  0.00  &        &         &        &        &          &        \\
model: rh                      &   5.32 &   0.00  &        &       &         &       &          &        & 61.05  &  0.00  &        &         &        &        &          &        \\
gcc                            &        &         &  -9.52 &  0.00 &         &       &   13.50  &   0.00 &        &        &   8.54 &    0.00 &        &        &   13.12  &   0.00 \\
acc                            &        &         &  14.49 &  0.00 &         &       &   10.29  &   0.00 &        &        &  54.08 &    0.00 &        &        &   35.06  &   0.00 \\
md                             &        &         &  -0.48 &  0.00 &         &       &    2.42  &   0.00 &        &        &  -0.19 &    0.00 &        &        &   -3.79  & - 0.00 \\
dc                             &        &         &   0.00 &  0.99 &         &       &    5.56  &   0.00 &        &        &   2.99 &    0.00 &        &        &    1.38  &   0.00 \\
ecomms                         &        &         &   0.07 &  0.00 &         &       &   -0.04  &   0.00 &        &        &   0.08 &    0.00 &        &        &    2.39  &   0.00 \\
ci                             &        &         &  10.79 &  0.00 &         &       &   -6.34  &   0.03 &        &        & -36.41 &    0.00 &        &        &    0.23  & - 0.00 \\
div                            &        &         &  -1.95 &  0.00 &         &       &    2.90  &   0.00 &        &        &  -4.98 &    0.00 &        &        & -126.05  & - 0.00 \\
cscore                         &        &         &        &       &         &       &   26.88  &   0.00 &        &        &        &         &        &        &   -6.44  &   0.00 \\
beta                           &        &         &        &       &  -2.25  &  0.00 &          &        &        &        &        &         &  16.76 &   0.00 &   13.09  &        \\
linearity                      &        &         &        &       & -49.31  &  0.00 &          &        &        &        &        &         & 111.74 &   0.00 &          &        \\
has\_temp                      &        &         &        &       &  57.31  &  0.00 &          &        &        &        &        &         & -49.15 &   0.00 &          &        \\ \hline
has\_temp $\times$ temp        &        &         &        &       & -10.93  &  0.00 &          &        &        &        &        &         & -46.62 &   0.00 &          &        \\ \hline
N\_obs                         &\multicolumn{2}{r|}{22400}&\multicolumn{2}{r|}{22400}&\multicolumn{2}{r|}{22400}&\multicolumn{2}{r|}{19200}   &\multicolumn{2}{r|}{22400}&\multicolumn{2}{r|}{22400}&\multicolumn{2}{r|}{22400}&\multicolumn{2}{r|}{19200}  \\ 
R2 (R2 adjusted)               &\multicolumn{2}{r|}{0.59 (0.59)}&\multicolumn{2}{r|}{0.60 (0.60)}&\multicolumn{2}{r|}{0.60 (0.60)}&\multicolumn{2}{r|}{0.72 (0.72)} &\multicolumn{2}{r|}{0.64 (0.64)}&\multicolumn{2}{r|}{0.94 (0.94)}&\multicolumn{2}{r|}{0.86 (0.86)}&\multicolumn{2}{r|}{0.93 (0.93)} \\
F statistic (p-value)          &\multicolumn{2}{r|}{2919.90 (0.00)}&\multicolumn{2}{r|}{2558.70 (0.00)}&\multicolumn{2}{r|}{3320.73 (0.00)}&\multicolumn{2}{r|}{3552.92 (0.00)} &\multicolumn{2}{r|}{3569.87 (0.00)}&\multicolumn{2}{r|}{25276.34 (0.00)}&\multicolumn{2}{r|}{14216.83 (0.00)}&\multicolumn{2}{r|}{18544.79 (0.00)} \\ \hline
\end{tabular}
}
\caption{Factors explaining the difference between results of HyperFast against LLP+WG or trivial approach. OLS regressions. \label{tab:regression}}
\end{table*}

Based on the data in Table \ref{tab:regression}, we notice that the network type significantly affects HyperFast performance. Embedder choice significantly impacts the outcomes: HMCS yielding on average results better by about 11 percent points than the results from BFKL. If model information is unavailable, the network characteristics serve as effective proxies. Performance correlates with clustering coefficients and the community structure of the graph. Embedder that successfully retrieve angular coordinates (a proxy for similarity space) achieve better results versus SOTA. Overall, the usage of HyperFast improves the compression results against trivial. The models explain a significant share of variability when explaining the comparison against trivial method; $R^2$ values for SOTA comparisons are reasonably high, given sample sizes.

\section{Adjustments of HyperFast}\label{sec:furthopt}

In this section we provide the details of various adjustments that could be applied to HyperFast to hopefully improve its compression quality and time.

\paragraph{Step 1: optimize distance buckets.} In Section \ref{sec:hyperfast} we have proposed to split the interval of possible node distances into intervals of equal length.
While this method works quite well in practice, unequal lengths may provide better results with less buckets. Therefore, we propose the following improvements. We first compute the
values of $m_c$ and $m'_c$ for $2B_d$ small buckets. When these values are known, we combine these small buckets into $B_d/2$ larger buckets in an optimal way. It is possible
to determine the size of the compressed stream for a given set of large buckets using entropy formulas, and find the optimal partition using a Dynamic Programming algorithm running
in negligible time $O(B_d^3)$. This doubles the running time of computing $m'_c$ while halving the running time of the actual compression part. Since the actual compression part tends
to take more time, this actually improves the total running time while hopefully improving the compression.

\paragraph{Step 2: directed degree-based radial distances.} CLOVE, LPCS an HMCS sort all the vertices by decreasing degree, and a vertex of index $i$
in this order gets radial distance $2\beta\log(i)+2(1-\beta)\log(N)$. This formula is based on the PSO model \cite{papa}. The value $\beta$ is taken to be $\frac{1}{\gamma-1}$,
where $\gamma$ is the estimated power-law exponent $\gamma$ of the vertex distribution. If the graph follows the power law degree distribution exactly, we obtain that the
radial distance of a vertex of degree $d$ is close to $2\log(N) - 2\log(d)$.

Real-world graphs may not exactly follow the power law distribution, especially for the low-degree vertices, which are the majority. Therefore it makes sense to directly use
radial distance $2\log(N) - 2\log(d)$ instead of the one provided by the embedder. This is especially useful for directed graphs, in which we can split every node into
two (in-node and out-node) and set their radial distances respectively based on the in-degree and out-degree, in particular, nodes of in-degree or out-degree 0 having radial distance
of infinity and thus essentially not being considered as possible sources or targets. Thus, we essentially follow the ideas outlined in Subection \ref{sec:directed}.
One difference is that the angular position for matching in- and out-nodes is always the same, which makes sense for compression purposes, as otherwise we would need to compress the information
about correlations between the two orders.

\paragraph{Step 3: optimize radial buckets.} In Step 1 we have optimized distance buckets, so we can also think about optimizing radial buckets. While the method we used for
distance buckets does not work here, we can use a simpler, approximate one. Assume that we compress the graph based only on the information of the degree of the source node.
If there are $q$ nodes of degree $d$, entropy coding their outgoing edges would take us $qN_{\rm{out}} H(\frac{d}{N_{\rm{out}}})$ bits, where $N_{\rm{out}}$ is the number of nodes
of out-degree greater than 0. Essentially, we get a separate radial bucket for every possible degree; we then use a Dynamic Programming method similar to the one from Step 1 to
combine them into $B_r$ optimal distance buckets.

\paragraph{Step 4: triangle-based edge prediction.} So far, we have been estimating the probability of nodes $u$ and $v$ being connected solely on their bucketed distance $|uv|$. However,
note that the HyperFast algorithm can actually freely access the bucketed distances of both nodes $u$ and $v$ from the center. Therefore, instead of computing the values of $m_c$ and
$m'_c$ for every distance bucket $c$, we compute values $m_{c,b,b'}$ and $m'_{c,b,b'}$ for every distance bucket $c$ and for every radial bucket of both in-node and out-node.
Probabilities $\frac{m_{c,b,b'}}{m'_{c,b,b'}}$ may let us improve the quality of entropy coding over using simply $\frac{m_c}{m'_c}$, especially if hyperbolic geometry is
not that suitable for modelling the given network. This comes at the cost of having to
include a bigger table in the compressed stream, however, this is negligible for larger graphs.

\paragraph{Step 5: local search.} Yet another possible idea is to, for every node, try to move it into a different radial buckets rather than the one computed from its degree,
see if it would improve the compression ratio, and actually move it if it does. (This is similar to the local search from \cite{dhrg_sea}.)

To test these five ideas, we have run preliminary experiments on real-world graphs. In Table \ref{adjsum}, for each embedder, each real-world graph,
and each number $i$ from 0 to 5, we present the ratio of the HyperFast compressed file to the file compressed using LLP+WebGraph when performing steps from 1 to $i$.
We also present the running time of HyperFast.

All steps 1-4 substantially improve the compression for some graphs, while not substantially worsening
the compression for others, and not worsening the running time. Step 5 substantially improves the compression for only one graph ({\sc uk-2002}) while greatly worsening the running time,
so we have decided not to use it.

\begin{table}
\resizebox{\columnwidth}{!}{
\begin{tabular}{|l|rrrrrr|} \hline
embedder                      & \multicolumn{6}{c|}{HMCS}\\ \hline 
number of steps               & 0& 1& 2& 3& 4& 5\\ \hline 
hepph                         & $(135.34\pm0.78)\%$& $(136.68\pm0.75)\%$& $(127.21\pm0.82)\%$& $(126.98\pm0.83)\%$& $(105.63\pm0.44)\%$& $(101.86\pm0.53)\%$ \\ 
 $90028.0\pm104.8$B & $0.38\pm0.01s$& $0.34\pm0.01s$& $0.39\pm0.01s$& $0.36\pm0.01s$& $0.36\pm0.01s$& $0.79\pm0.02s$ \\ \hline 
astroph                       & $(127.82\pm0.21)\%$& $(128.51\pm0.26)\%$& $(114.83\pm0.19)\%$& $(114.64\pm0.23)\%$& $(108.08\pm0.17)\%$& $(106.62\pm0.20)\%$ \\ 
 $249541.0\pm285.0$B & $0.74\pm0.02s$& $0.68\pm0.01s$& $0.76\pm0.02s$& $0.67\pm0.01s$& $0.66\pm0.01s$& $1.40\pm0.03s$ \\ \hline 
enron-nat                     & $(118.29\pm0.45)\%$& $(118.92\pm0.52)\%$& $(110.65\pm0.44)\%$& $(106.30\pm0.38)\%$& $(97.28\pm0.40)\%$& $(94.21\pm0.39)\%$ \\ 
 $201694.2\pm877.3$B & $0.76\pm0.02s$& $0.69\pm0.02s$& $0.69\pm0.01s$& $0.66\pm0.01s$& $0.66\pm0.01s$& $2.04\pm0.05s$ \\ \hline 
facebook                      & $(117.14\pm0.66)\%$& $(117.64\pm0.62)\%$& $(100.70\pm0.93)\%$& $(101.33\pm0.92)\%$& $(89.23\pm0.84)\%$& $(86.72\pm0.76)\%$ \\ 
 $69308.4\pm215.6$B & $0.23\pm0.00s$& $0.22\pm0.00s$& $0.21\pm0.00s$& $0.22\pm0.01s$& $0.22\pm0.00s$& $0.45\pm0.01s$ \\ \hline 
google                        & $(101.06\pm0.11)\%$& $(98.58\pm0.10)\%$& $(99.30\pm0.14)\%$& $(93.74\pm0.21)\%$& $(87.28\pm0.16)\%$& $(83.93\pm0.35)\%$ \\ 
 $2972264.2\pm1570.1$B & $20.00\pm0.35s$& $17.03\pm0.23s$& $17.24\pm0.28s$& $15.07\pm0.17s$& $15.07\pm0.17s$& $39.62\pm0.35s$ \\ \hline 
metroidvania-5                & $(99.34\pm0.74)\%$& $(99.62\pm0.74)\%$& $(90.62\pm0.71)\%$& $(87.57\pm0.69)\%$& $(83.19\pm0.69)\%$& $(83.05\pm0.67)\%$ \\ 
 $153614.8\pm1363.6$B & $0.37\pm0.01s$& $0.35\pm0.01s$& $0.33\pm0.00s$& $0.34\pm0.00s$& $0.34\pm0.00s$& $0.85\pm0.02s$ \\ \hline 
condmat                       & $(91.33\pm0.19)\%$& $(91.57\pm0.14)\%$& $(92.15\pm0.20)\%$& $(91.52\pm0.30)\%$& $(85.58\pm0.18)\%$& $(85.27\pm0.30)\%$ \\ 
 $149603.2\pm152.8$B & $0.55\pm0.01s$& $0.47\pm0.01s$& $0.53\pm0.01s$& $0.41\pm0.01s$& $0.40\pm0.01s$& $0.98\pm0.03s$ \\ \hline 
slashdot                      & $(90.87\pm0.42)\%$& $(91.10\pm0.39)\%$& $(77.71\pm0.34)\%$& $(76.65\pm0.35)\%$& $(71.17\pm0.31)\%$& $(70.91\pm0.32)\%$ \\ 
 $1209621.8\pm5347.8$B & $2.51\pm0.05s$& $2.31\pm0.05s$& $2.53\pm0.05s$& $2.29\pm0.05s$& $2.27\pm0.05s$& $4.70\pm0.10s$ \\ \hline 
comedy-full                   & $(89.46\pm0.19)\%$& $(90.21\pm0.23)\%$& $(85.91\pm0.21)\%$& $(83.76\pm0.28)\%$& $(78.68\pm0.22)\%$& $(78.09\pm0.21)\%$ \\ 
 $628558.1\pm790.3$B & $2.24\pm0.08s$& $1.93\pm0.06s$& $2.10\pm0.06s$& $1.64\pm0.05s$& $1.60\pm0.05s$& $3.77\pm0.11s$ \\ \hline 
comedy                        & $(89.15\pm0.26)\%$& $(90.00\pm0.28)\%$& $(84.76\pm0.22)\%$& $(83.83\pm0.20)\%$& $(80.20\pm0.20)\%$& $(79.73\pm0.28)\%$ \\ 
 $578086.1\pm791.3$B & $1.76\pm0.03s$& $1.53\pm0.02s$& $1.67\pm0.03s$& $1.34\pm0.02s$& $1.32\pm0.02s$& $3.03\pm0.05s$ \\ \hline 
Drosophila2                   & $(86.85\pm0.33)\%$& $(87.12\pm0.37)\%$& $(71.37\pm0.28)\%$& $(70.60\pm0.31)\%$& $(68.95\pm0.29)\%$& $(68.84\pm0.26)\%$ \\ 
 $16362.8\pm44.4$B & $0.07\pm0.00s$& $0.07\pm0.00s$& $0.06\pm0.00s$& $0.06\pm0.00s$& $0.06\pm0.00s$& $0.13\pm0.00s$ \\ \hline 
Mouse3                        & $(88.74\pm0.87)\%$& $(88.55\pm0.91)\%$& $(88.46\pm0.94)\%$& $(88.13\pm0.96)\%$& $(84.93\pm0.72)\%$& $(84.39\pm0.77)\%$ \\ 
 $69472.3\pm163.7$B & $0.20\pm0.00s$& $0.19\pm0.00s$& $0.19\pm0.00s$& $0.25\pm0.00s$& $0.25\pm0.00s$& $0.43\pm0.01s$ \\ \hline 
enron                         & $(88.42\pm0.30)\%$& $(88.72\pm0.34)\%$& $(87.94\pm0.39)\%$& $(87.18\pm0.41)\%$& $(81.87\pm0.38)\%$& $(81.28\pm0.35)\%$ \\ 
 $313475.9\pm954.4$B & $0.93\pm0.02s$& $0.84\pm0.01s$& $0.92\pm0.02s$& $0.83\pm0.01s$& $0.83\pm0.01s$& $1.84\pm0.03s$ \\ \hline 
openflights                   & $(88.18\pm0.42)\%$& $(88.47\pm0.49)\%$& $(81.56\pm0.50)\%$& $(81.08\pm0.52)\%$& $(79.29\pm0.51)\%$& $(78.22\pm0.48)\%$ \\ 
 $23531.6\pm64.4$B & $0.12\pm0.01s$& $0.10\pm0.00s$& $0.10\pm0.00s$& $0.10\pm0.00s$& $0.10\pm0.00s$& $0.22\pm0.00s$ \\ \hline 
grqc                          & $(87.07\pm0.58)\%$& $(86.81\pm0.60)\%$& $(81.47\pm0.64)\%$& $(80.06\pm0.58)\%$& $(72.82\pm0.43)\%$& $(71.78\pm0.51)\%$ \\ 
 $17719.6\pm39.6$B & $0.11\pm0.00s$& $0.09\pm0.00s$& $0.10\pm0.00s$& $0.08\pm0.00s$& $0.08\pm0.00s$& $0.19\pm0.00s$ \\ \hline 
wordassociation-2011          & $(79.25\pm0.10)\%$& $(79.36\pm0.10)\%$& $(75.33\pm0.11)\%$& $(72.88\pm0.10)\%$& $(71.37\pm0.08)\%$& $(71.41\pm0.15)\%$ \\ 
 $95342.0\pm129.2$B & $0.22\pm0.00s$& $0.21\pm0.00s$& $0.20\pm0.00s$& $0.18\pm0.00s$& $0.18\pm0.00s$& $0.44\pm0.01s$ \\ \hline 
oregon                        & $(68.84\pm0.90)\%$& $(68.85\pm0.80)\%$& $(65.76\pm0.79)\%$& $(63.35\pm0.77)\%$& $(59.67\pm0.69)\%$& $(58.21\pm0.69)\%$ \\ 
 $42102.8\pm405.2$B & $0.24\pm0.01s$& $0.20\pm0.00s$& $0.19\pm0.00s$& $0.16\pm0.00s$& $0.16\pm0.00s$& $0.38\pm0.01s$ \\ \hline 
ias                           & $(64.12\pm0.35)\%$& $(64.30\pm0.36)\%$& $(64.08\pm0.32)\%$& $(62.31\pm0.31)\%$& $(58.40\pm0.34)\%$& $(57.64\pm0.36)\%$ \\ 
 $113834.4\pm600.8$B & $0.51\pm0.01s$& $0.43\pm0.01s$& $0.43\pm0.01s$& $0.36\pm0.00s$& $0.36\pm0.01s$& $0.89\pm0.01s$ \\ \hline 
amazon                        & $(64.29\pm0.14)\%$& $(64.37\pm0.11)\%$& $(65.15\pm0.05)\%$& $(64.09\pm0.10)\%$& $(62.99\pm0.06)\%$& $(62.44\pm0.12)\%$ \\ 
 $1904184.4\pm751.6$B & $8.31\pm0.13s$& $6.65\pm0.11s$& $7.54\pm0.11s$& $5.12\pm0.08s$& $5.11\pm0.07s$& $12.53\pm0.21s$ \\ \hline 
fork-10                       & $(79.72\pm0.07)\%$& $(79.83\pm0.07)\%$& $(79.95\pm0.07)\%$& $(79.57\pm0.07)\%$& $(75.28\pm0.06)\%$& $(75.05\pm0.06)\%$ \\ 
 $105321867.7\pm78003.3$B & $204.07\pm1.99s$& $204.07\pm1.43s$& $208.45\pm0.88s$& $210.00\pm0.84s$& $208.05\pm1.14s$& $351.74\pm14.54s$ \\ \hline 
patents                       & $(77.62\pm0.05)\%$& $(78.04\pm0.04)\%$& $(78.84\pm0.05)\%$& $(76.19\pm0.04)\%$& $(75.08\pm0.05)\%$& $(74.46\pm0.23)\%$ \\ 
 $29055809.3\pm12220.5$B & $103.13\pm3.55s$& $99.69\pm3.67s$& $107.59\pm3.98s$& $87.13\pm3.22s$& $87.20\pm3.23s$& $208.52\pm7.06s$ \\ \hline 
uk-2002                       & $(231.07\pm1.22)\%$& $(217.84\pm0.25)\%$& $(202.71\pm0.21)\%$& $(209.50\pm0.42)\%$& $(166.66\pm0.58)\%$& $(146.41\pm1.77)\%$ \\ 
 $65504894.8\pm25614.3$B & $574.39\pm11.57s$& $520.98\pm11.26s$& $514.56\pm11.74s$& $497.99\pm8.68s$& $495.56\pm8.88s$& $1196.46\pm23.96s$ \\ \hline 
enwiki-2013-hc                & $(83.56\pm0.19)\%$& $(83.63\pm0.20)\%$& $(82.65\pm0.20)\%$& $(82.23\pm0.20)\%$& $(80.74\pm0.19)\%$& $(80.34\pm0.16)\%$ \\ 
 $158983353.0\pm218718.4$B & $363.31\pm10.73s$& $353.86\pm10.75s$& $355.58\pm9.14s$& $346.18\pm7.36s$& $346.63\pm8.00s$& $605.42\pm13.61s$ \\ \hline 
\end{tabular}

}
\caption{The effects of adjustments on real-world graphs.
Bootstrapped 95\% confidence intervals for 
the ratio of the HyperFast compressed file to the file compressed using LLP+WebGraph (first column, under the name of the graph) and HyperFast running time,
when performing adjustment steps from 1 to $i$.
\label{adjsum}}
\end{table}

\section{Networks}\label{app:networks}
Table \ref{networklist} lists the real-world networks we used as benchmarks.

\def\citep#1{\cite{#1}}
\def\citet#1{\cite{#1}}
\begin{table*}[h!]
\centering
\scalebox{0.88}{
\begin{tabular}{lllrp{30em}} \toprule
name            & $|V|$& $|E|$ & type & source \\  \midrule

ias             & 23748&116828 & U & Internet autonomous systems \citep{bogu_internet} \\
facebook        & 4039 &176468 & U & Facebook social circles \citep{snapnets,tobias} \\
openflights     & 3397 &38460  & U & flight connections, OpenFlights website, \cite{mercatorembedding,hypclove}\\
grqc            &  4158&26844  & U & collaboration network: general relativity and quantum cosmology \citep{snapnets} \\
astroph         & 17903&393944 & U & collaboration network: astrophysics \citep{snapnets} \\
condmat         & 21363&182572 & U & collaboration network: condensed matter \citep{snapnets} \\
hepph           & 11204&235238 & U & collaboration network: high-energy physics \citep{snapnets} \\
Drosophila2     & 1770 &17810  & U & connectome, Drosophila optic medulla \citet{connectome_drosophila} \\ 
Mouse3          & 1076 &181622 & U & connectome, cells of mouse retina \citet{connectome_mouse23} \\ 
Human1          &  493 &15546  & U & connectome, areas of human brain \citet{connectome_human12} \\
amazon          &334863&1851744& U & Amazon co-purchase \cite{snapnets,tobias} \\
enron           &33696 &361622 & D & Enron e-mail network \cite{snapnets,rgb_bp} \\
enron-nat       &67251 &273649 & D & Enron e-mail network, WebGraph website \\
comedy          &62055 &629128 & U & co-starring in comedy movies: big component, IMDb \citep{papa} \\
comedy-full     &80468 &708306 & U & co-starring in comedy movies: all components, IMDb \citep{papa} \\
fork-10         &2480676&47860777& D & GitHub forking, vertices with degree $\geq 10$ \citep{euromed} \\
metroidvania-5  &24555&196728& B & Steam metroidvania review network$\rule{0cm}{0cm}^1$, vertices with degree $\geq 5$ \\
google          &855802&5066842& D & Google web graph \cite{snapnets,rgb_bp} \\
oregon          &11174 &46818  & U & Oregon autonomous systems \cite{snapnets,rgb_bp} \\
patents         &3764117&16511740& D & US patent citations \cite{snapnets,tobias} \\
slashdot        &77360&828161& D & Slashdot social network \cite{snapnets,tobias} \\
uk-2014-tpd     &1766010&1681750&D& UK top level domains, WebGraph website \\
uk-2002         &18459128&291989310&D& UK web graph, WebGraph website \\
enwiki-2013-hc  &4203294&101311589&D& English wikipedia, WebGraph website \\
wordassociation-2011&10617&72172&D& word association, WebGraph website \\
\bottomrule
\end{tabular}}
\caption{Our benchmark graphs, with original sources, and references to earlier work that used these graphs as benchmarks.
Type U means undirected, D means directed, B means bipartite.
$\rule{0cm}{0cm}^1$ Vertices are Steam users and Steam games tagged with 'metroidvania', edges exist if the given user has reviewed a given game.
\label{networklist}}
\end{table*}            

Table \ref{networklist} gives the list of networks we benchmark on. In all cases, we have removed self-loops (if they exist), 
restricted to only the giant component (it the network was not connected), and replaced undirected edges with pairs of directed edges (if the graph was undirected).
The restriction to the giant component was performed because the BFKL embedder assumes the graph to be connected.
In most cases, these changes do not significantly affect the compression, since self-loops are rare and almost all edges are in the giant component. The only
counterexample was the {\sc comedy} graph; we include both the connected and unconnected ({\sc comedy-full}) version of the graph, and for the experiments
using the BFKL embedder on {\sc comedy-full}, we add random extra edges to make the graph connected.
Note that while \cite{rgb_bp} lists \cite{snapnets} as the source of their graphs, the graph sizes do not match; we were unable to explain this discrepancy.


\section{Coefficients of variation and performance tables}\label{app:tables}

Table \ref{tab:cv} show the coefficients of variation
obtained in 10 runs of tested compression methods on real-world graphs.
We were unable to compute the results for CLOVE on large graphs due to the
high time complexity (CLOVE runs over a week on the Patents network).

Tables \ref{tab:rw:bpl} and \ref{tab:rw:time} show the compression quality 
obtained by the tested compression methods. Darker background means better result.

\begin{figure*}[h!]
\begin{center}
\includegraphics[width=\textwidth]{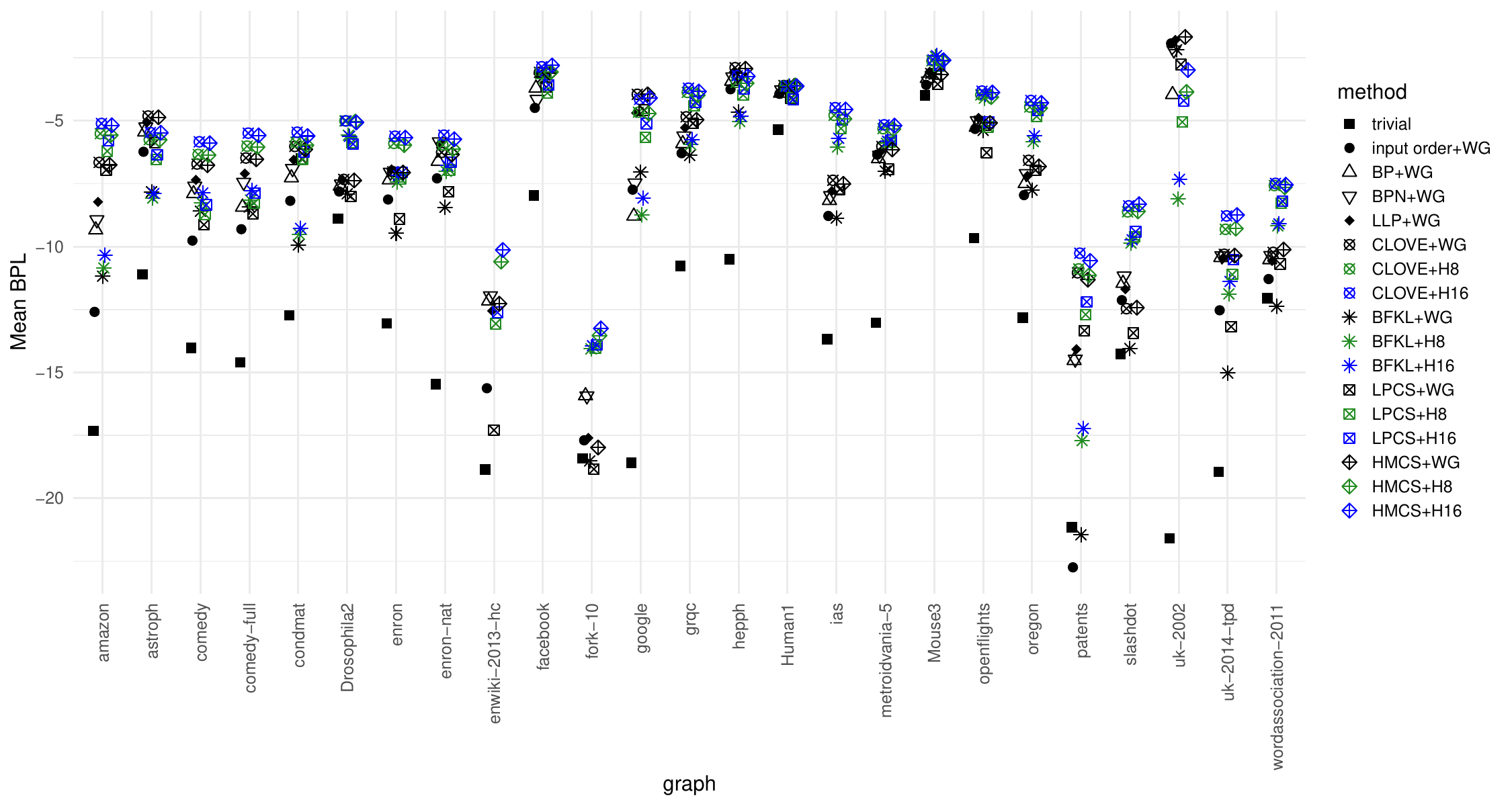}
\includegraphics[width=\textwidth]{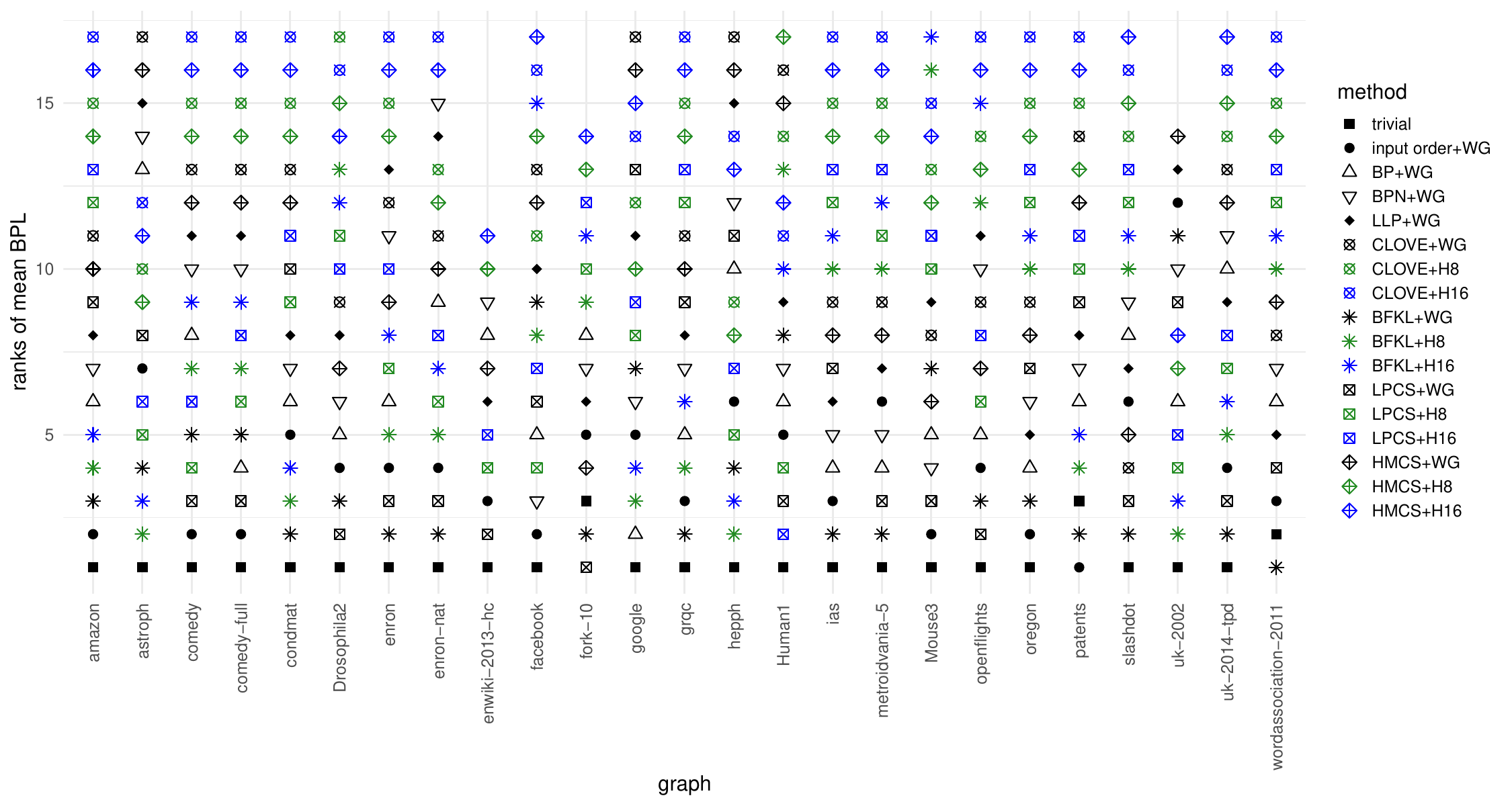}
\end{center}
\caption{Quality assessment of embedders on real-world graphs. The higher position indicates better quality. \label{qa_ranks1}}
\end{figure*}

\begin{figure*}[h!]
\begin{center}
\includegraphics[width=\textwidth]{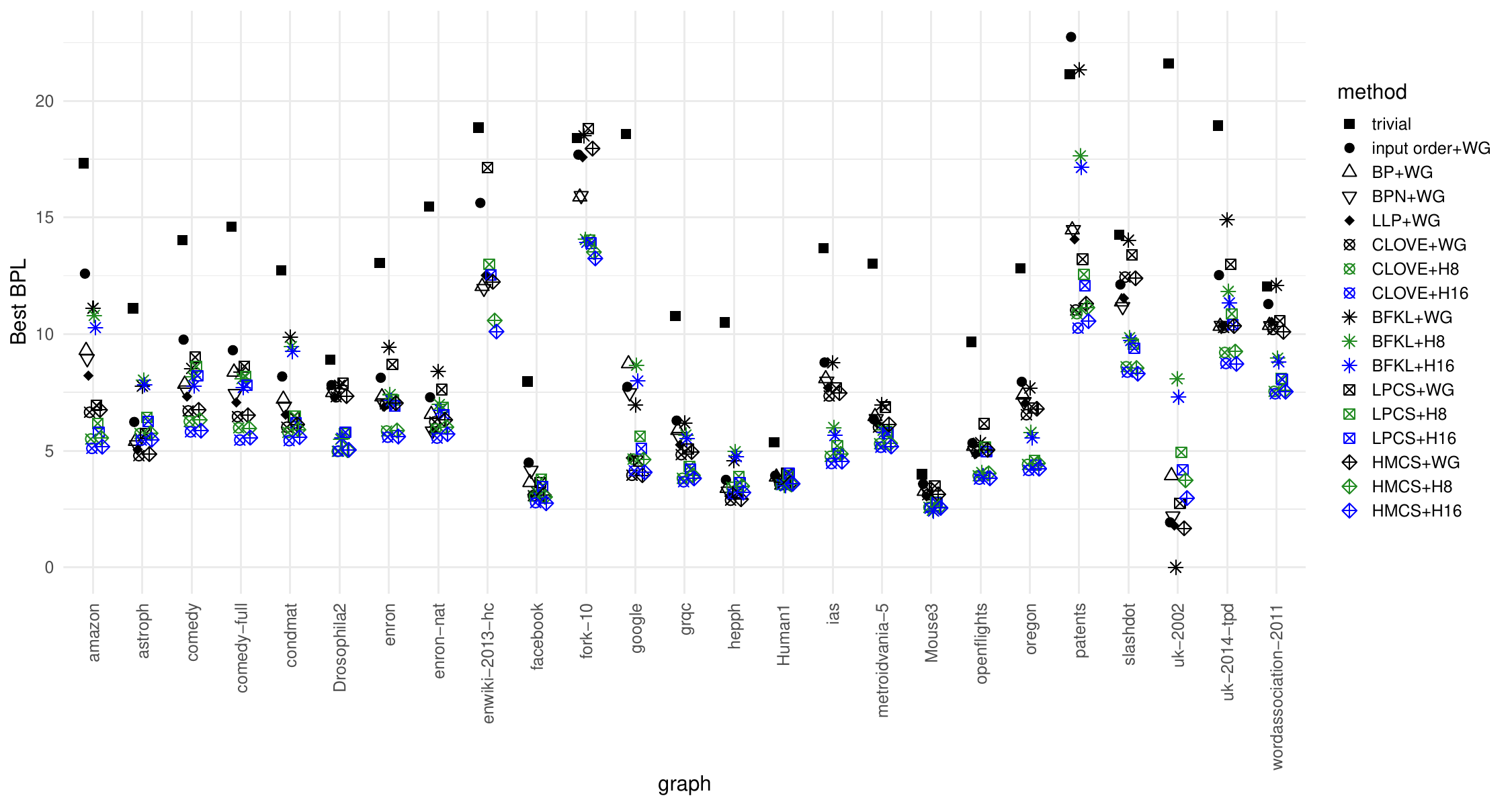}
\includegraphics[width=\textwidth]{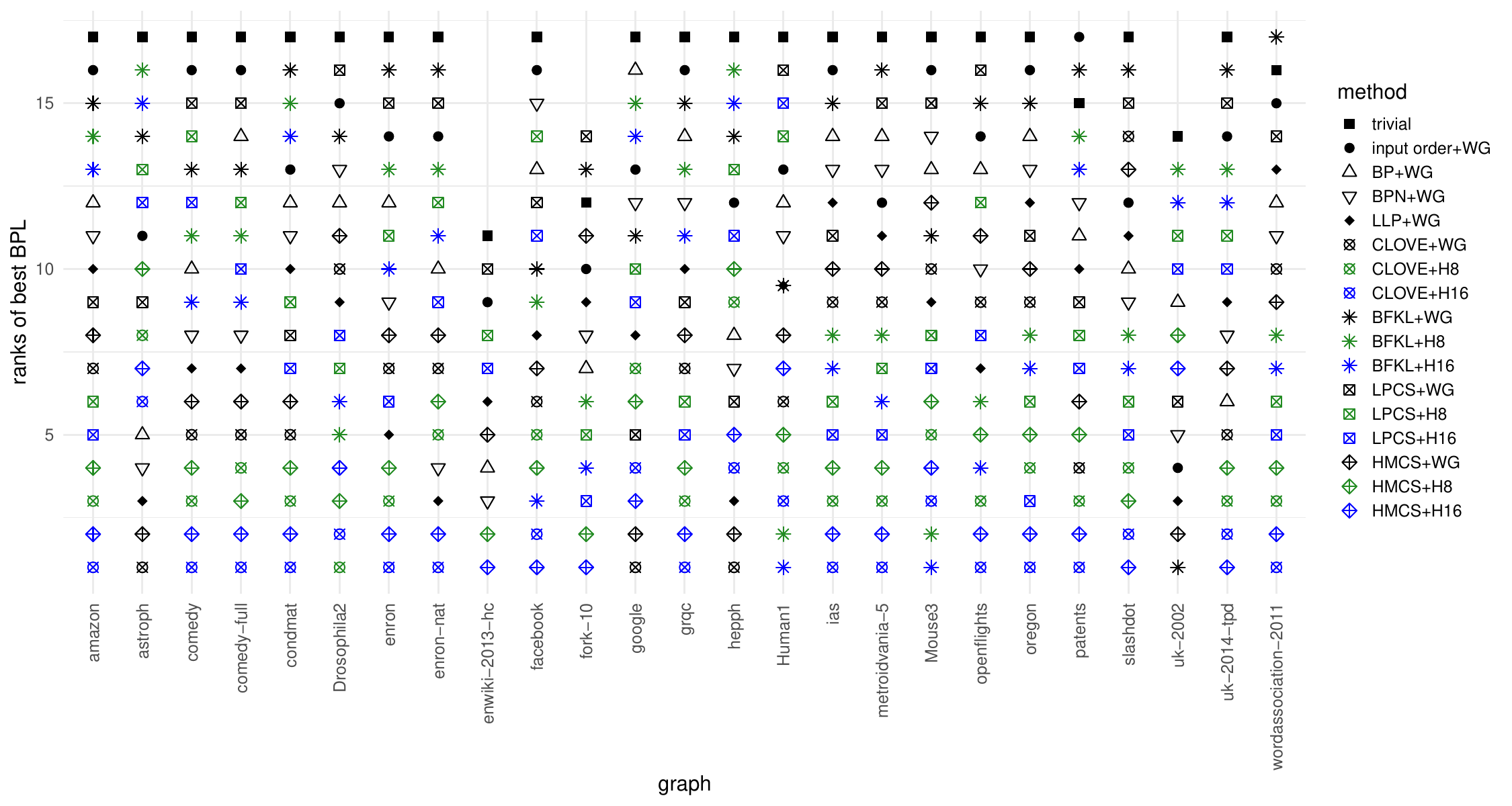}
\end{center}
\caption{Quality assessment of embedders on real-world graphs \label{qa_ranks2}}
\end{figure*}

In Figures \ref{qa_ranks1} and \ref{qa_ranks2} the compression quality is
presented as a graph.

\begin{figure*}[h!]
  \includegraphics[width=\textwidth]{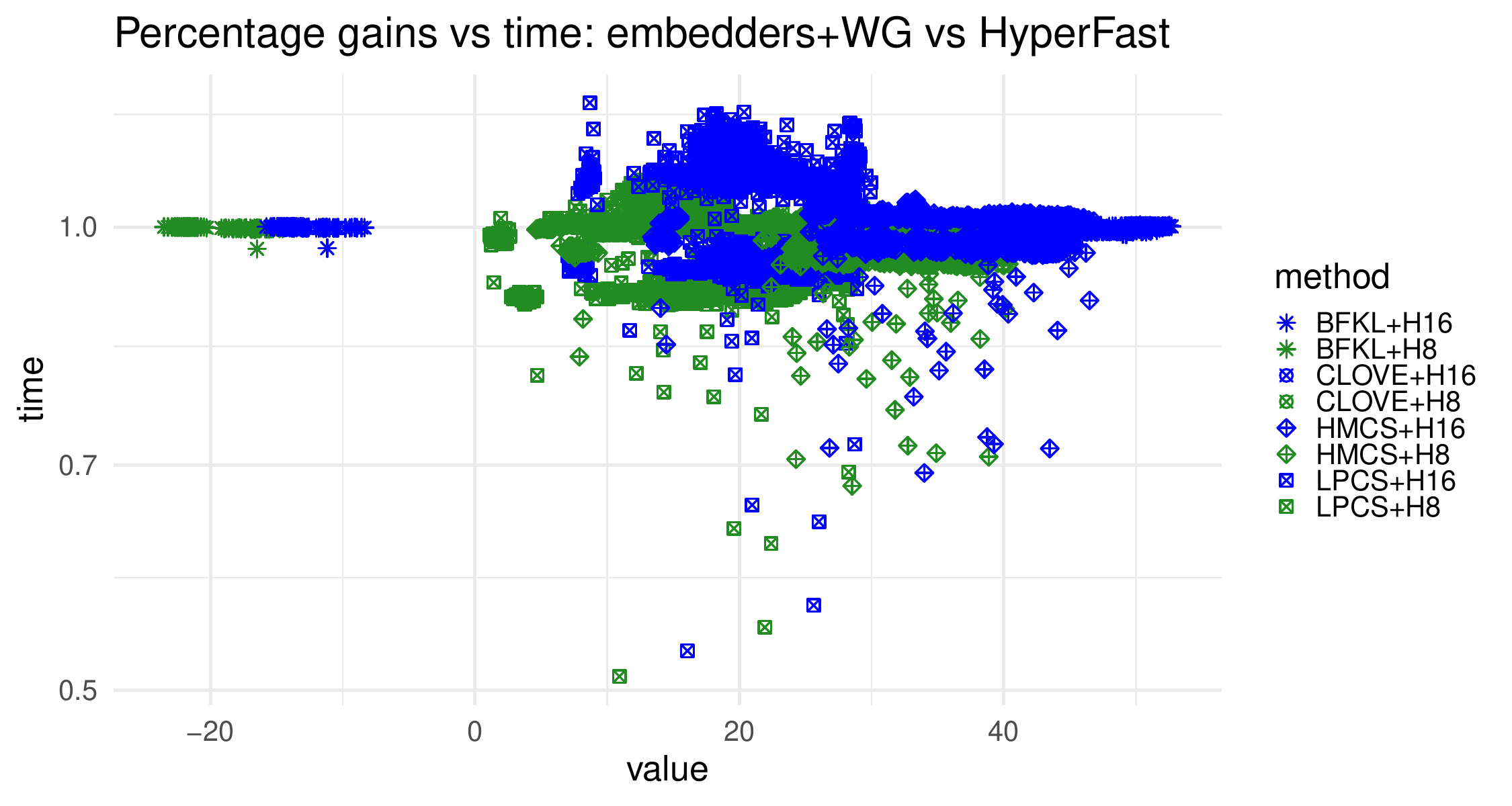}
  \caption{Percentage gains vs time: embedder+WebGraph vs HyperFast \label{time_gains_embs}}
\end{figure*}

\begin{figure*}[h!]
  \includegraphics[width=\textwidth]{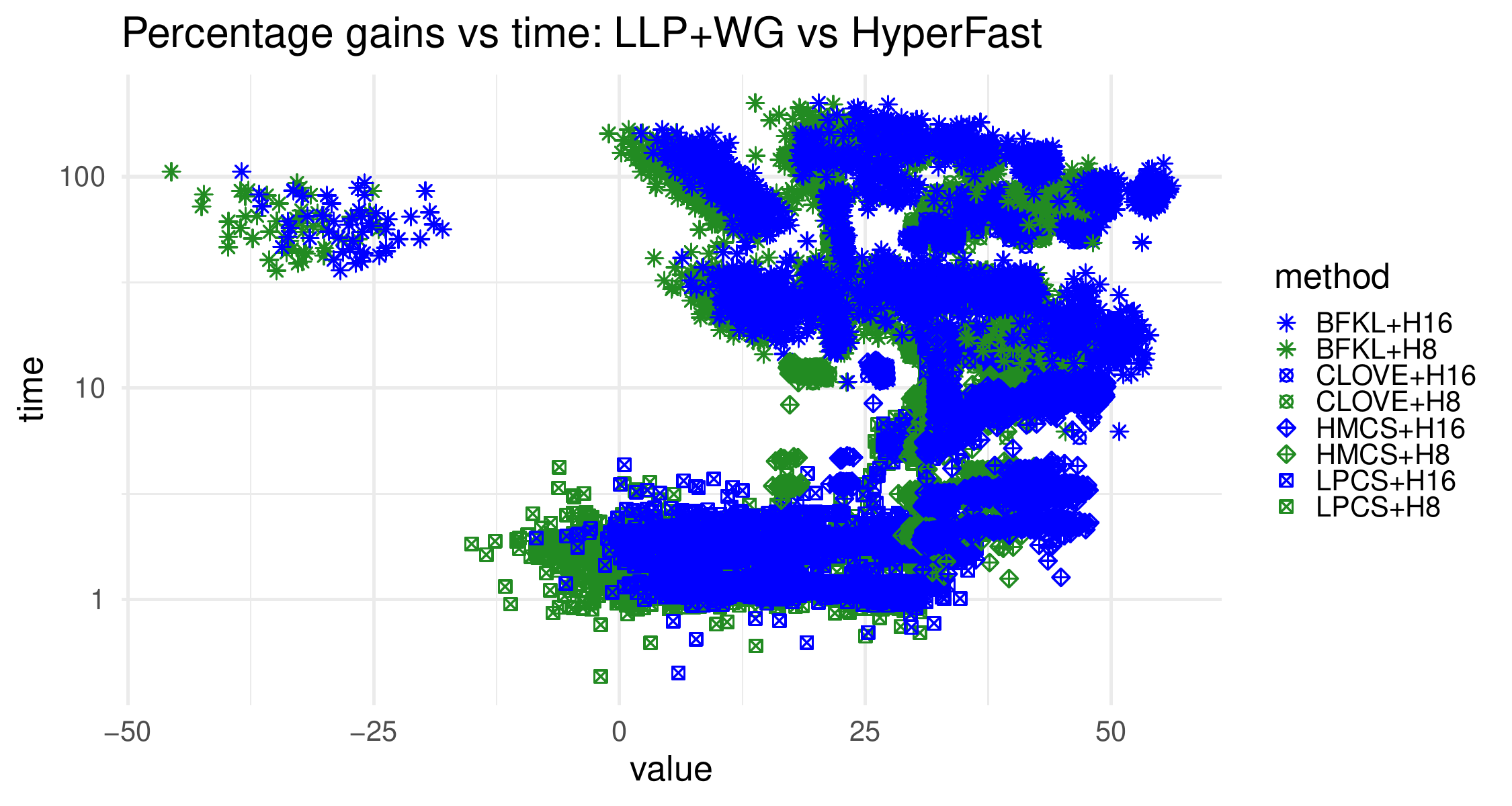}
  \caption{Percentage gains vs time: LLP+WebGraph vs HyperFast \label{time_gains_sota}}
\end{figure*}

In Figures \ref{time_gains_embs} and \ref{time_gains_sota} we compare
the percentage gains in the compression ratio versus the time required for
our embedders.

\section{Statistical inference}\label{app:inference}

\begin{figure*}[h!]
\begin{minipage}[b]{.5\columnwidth} \noindent
  \includegraphics[width=\textwidth]{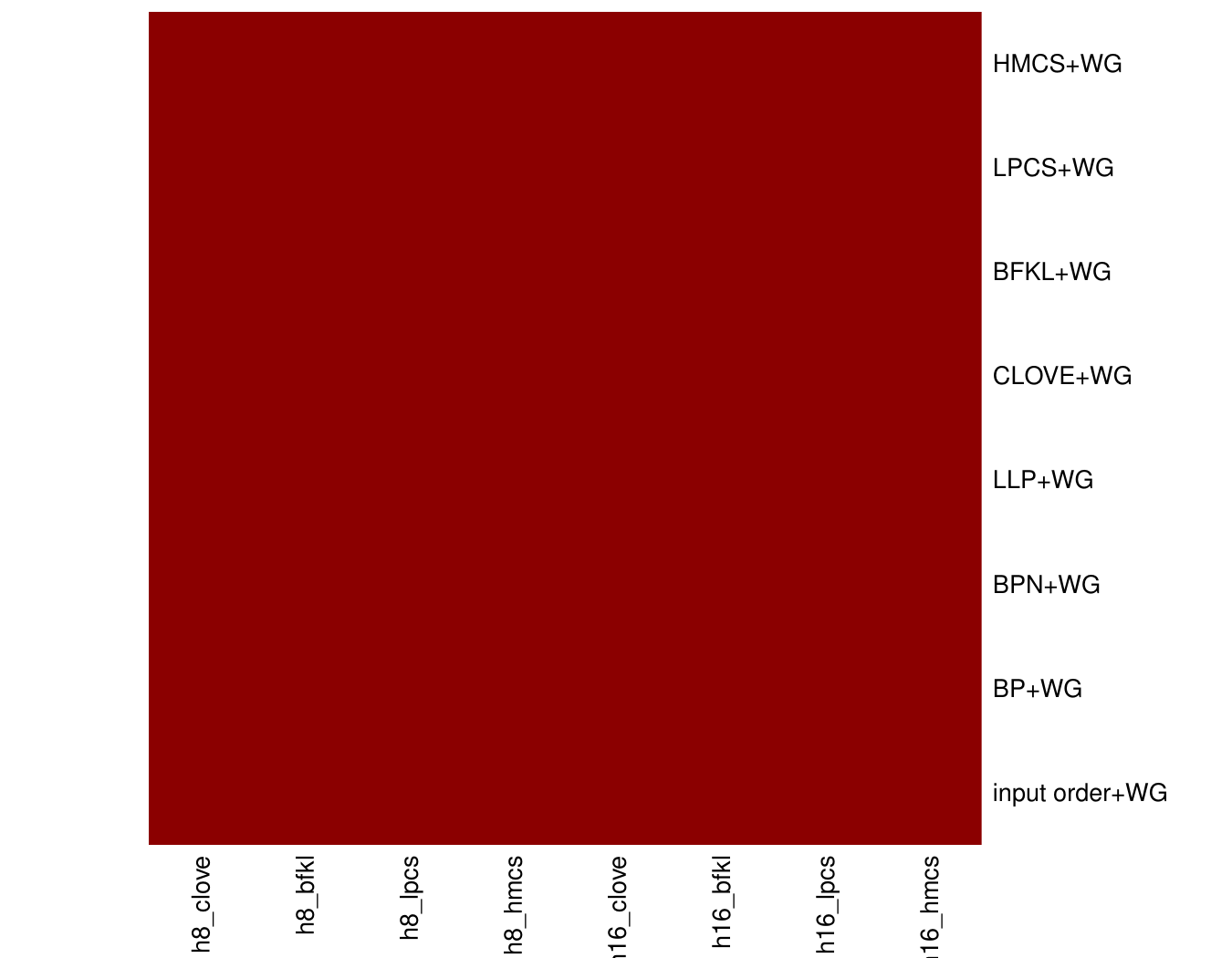}
  \subcaption{Full set}
\end{minipage}

\begin{minipage}[b]{.5\columnwidth} \noindent
  \includegraphics[width=\textwidth]{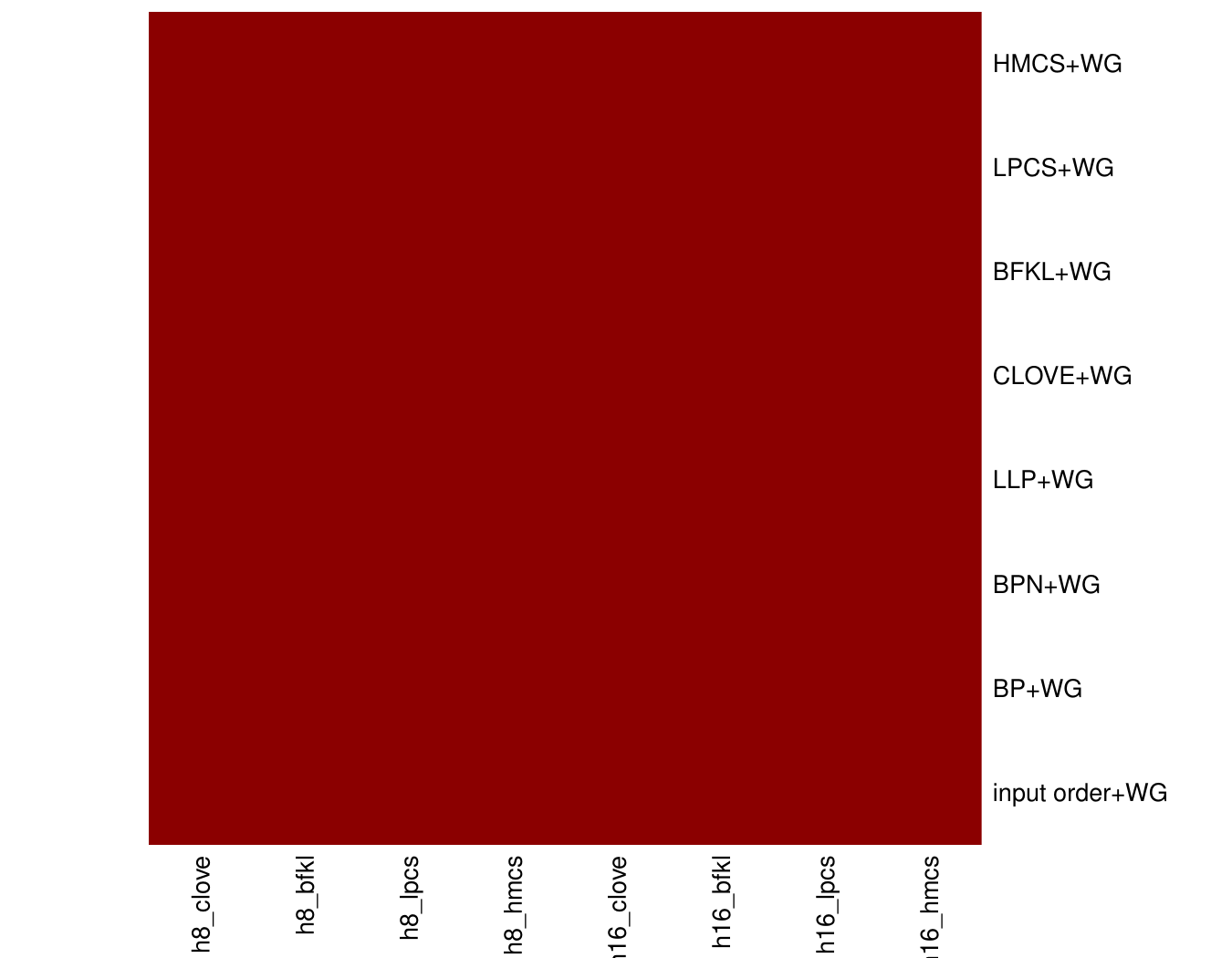}
  \subcaption{Non-geometric}
\end{minipage}
\begin{minipage}[b]{.5\columnwidth} \noindent
  \includegraphics[width=\textwidth]{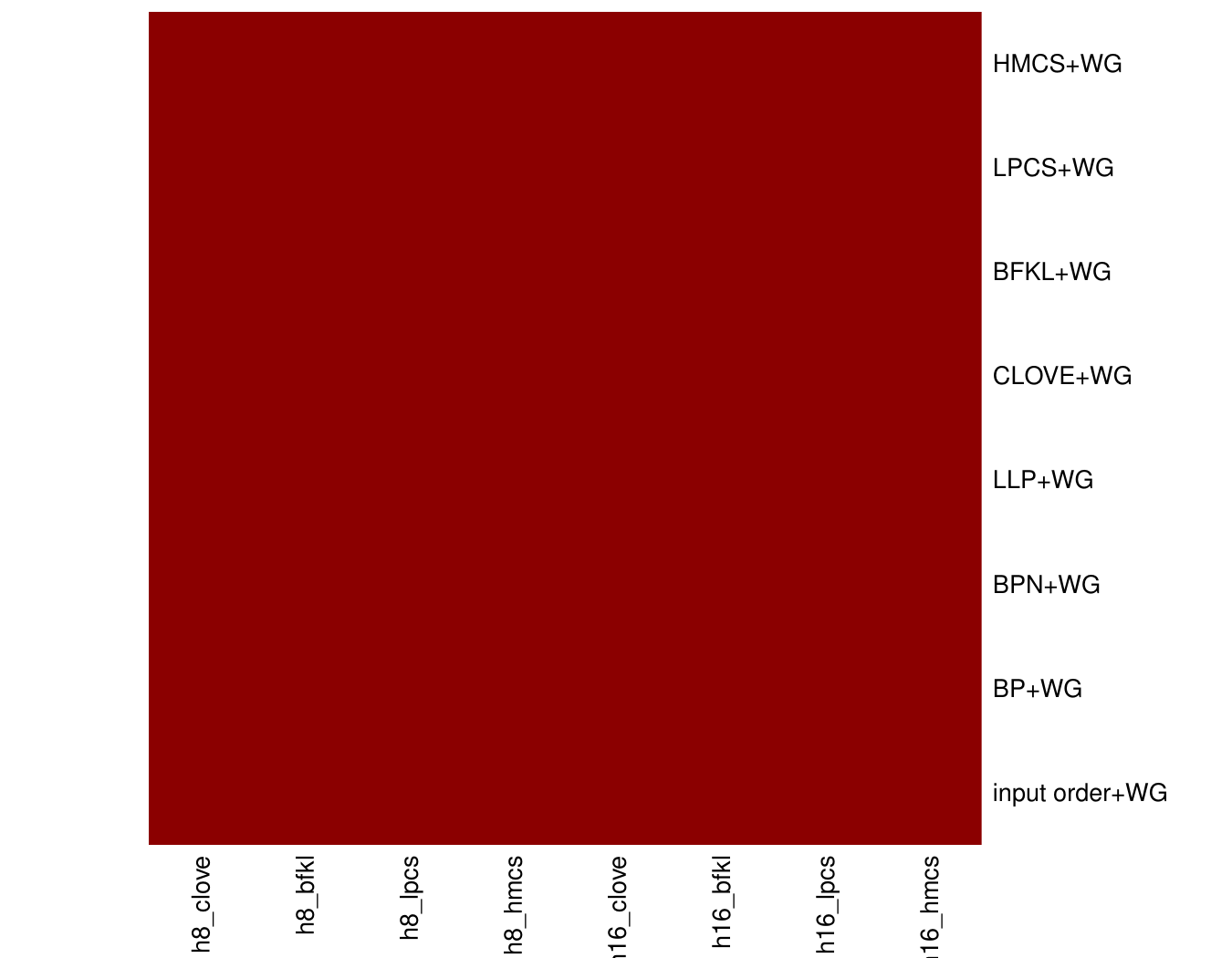}
  \subcaption{Large world}
\end{minipage}

\begin{minipage}[b]{.5\columnwidth} \noindent
  \includegraphics[width=\textwidth]{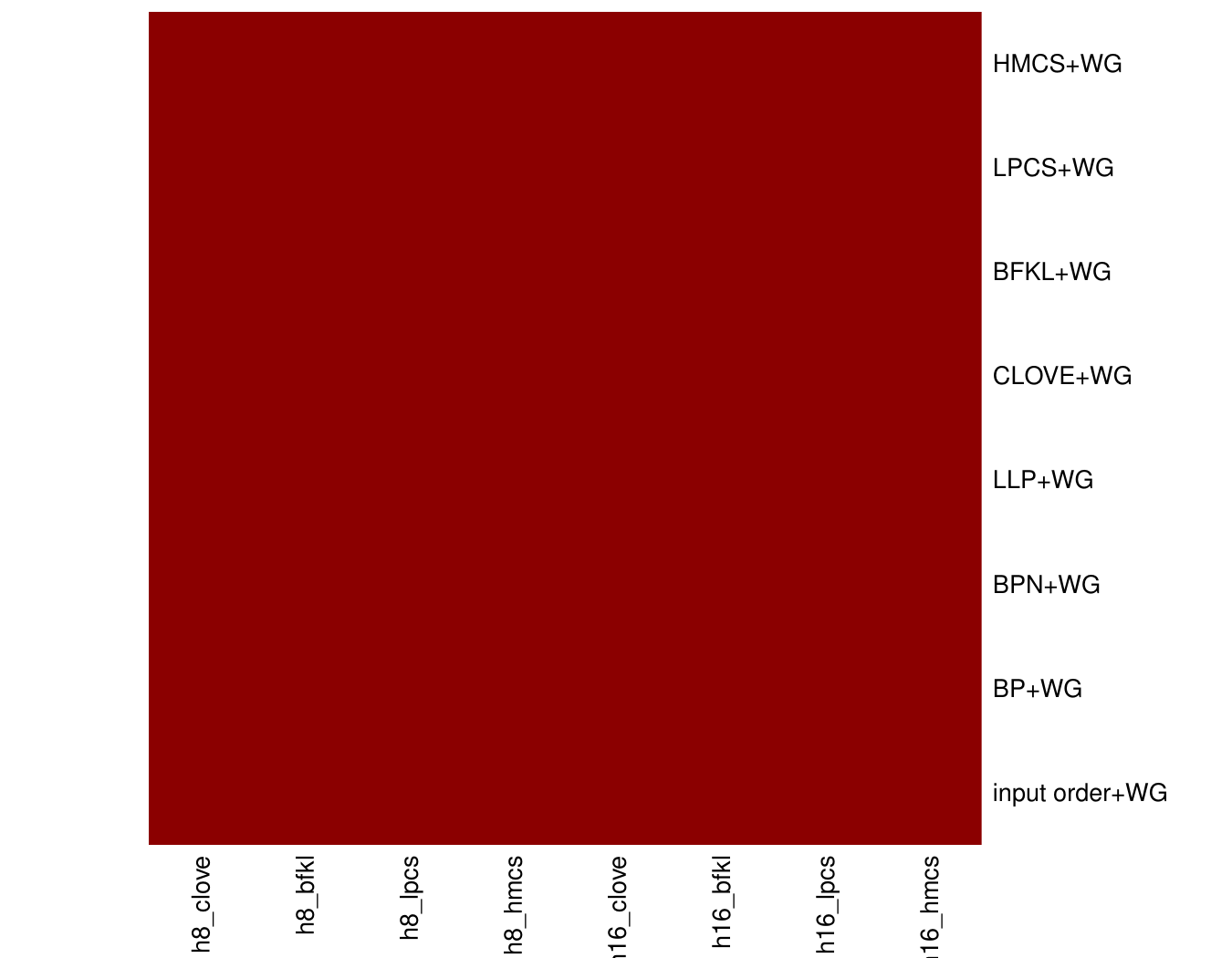}
  \subcaption{Small world}
\end{minipage}
\begin{minipage}[b]{.5\columnwidth} \noindent
  \includegraphics[width=\textwidth]{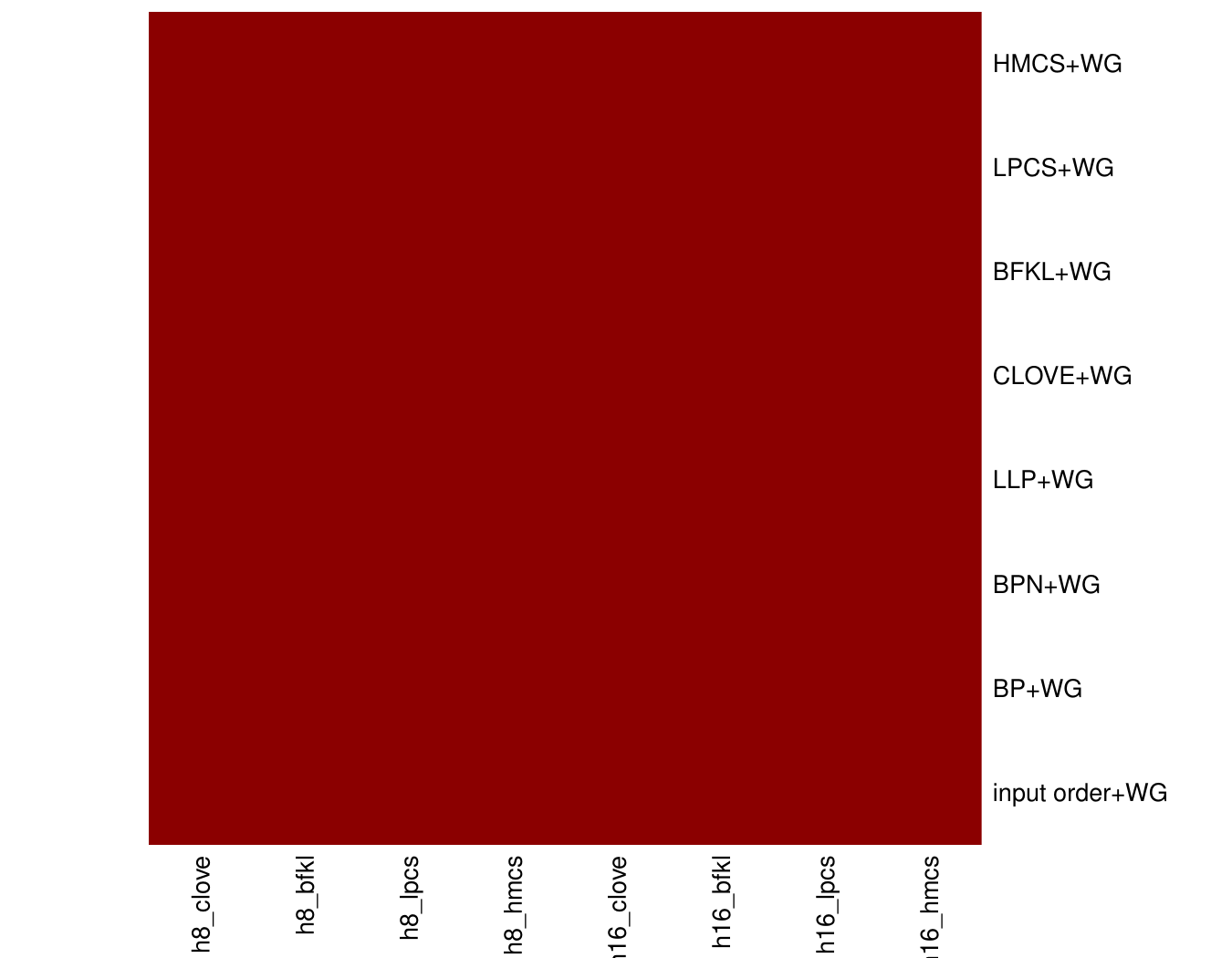}
  \subcaption{Ultra-small world}
\end{minipage}
  \caption{Comparison of the goodness of fit between results of HyperFast against WebGraph based methods. Red suggests that the HyperFast version yields better results and the difference is significant; orange suggests lack of significant difference, and yellow suggests significantly worse results for HyperFast. \label{fig:longlongsa}}
\end{figure*}

\begin{figure*}[h!]
\begin{minipage}[b]{.5\columnwidth} \noindent
  \includegraphics[width=\textwidth]{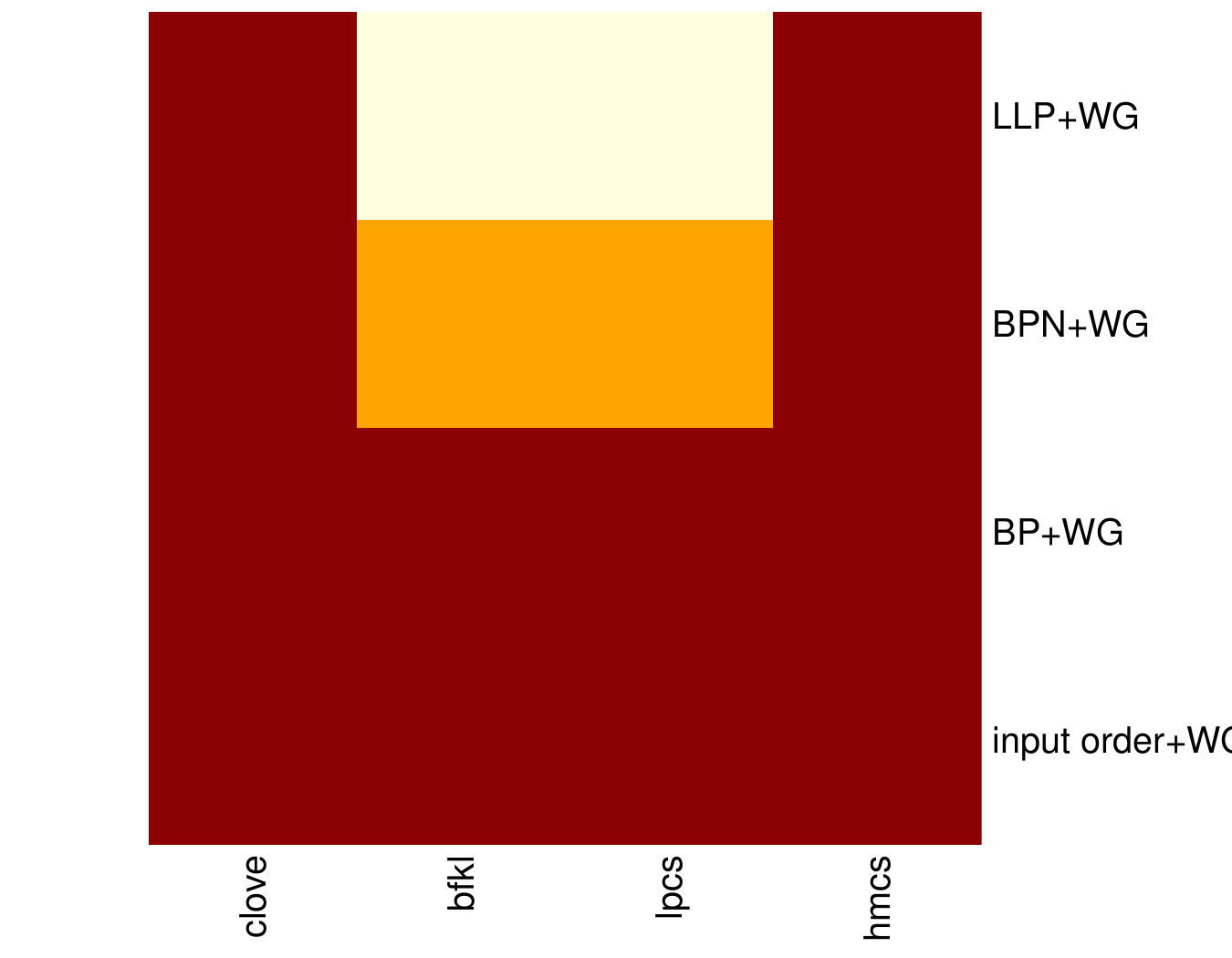}
  \subcaption{Full set}
\end{minipage}

\begin{minipage}[b]{.5\columnwidth} \noindent
  \includegraphics[width=\textwidth]{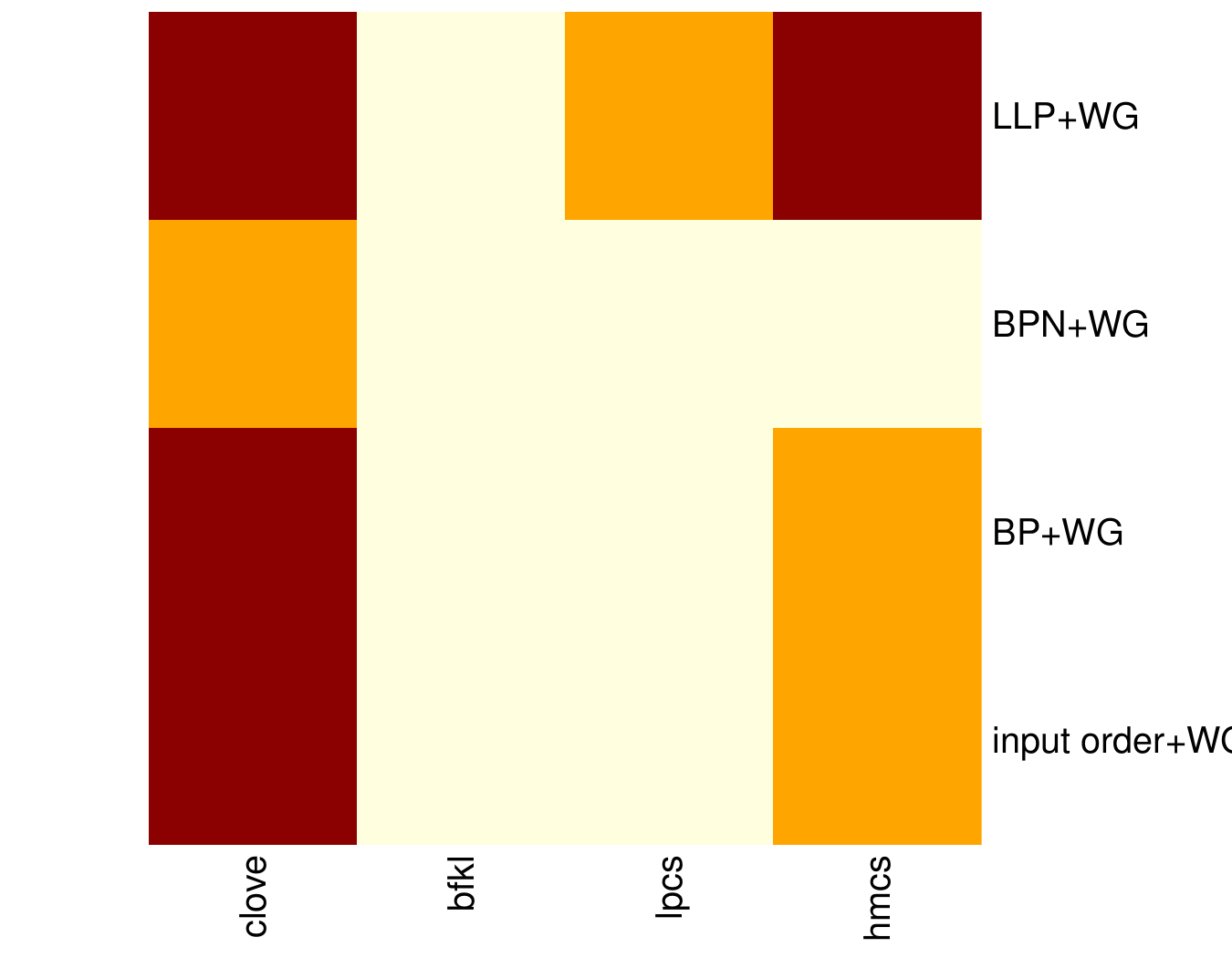}
  \subcaption{Non-geometric}
\end{minipage}
\begin{minipage}[b]{.5\columnwidth} \noindent
  \includegraphics[width=\textwidth]{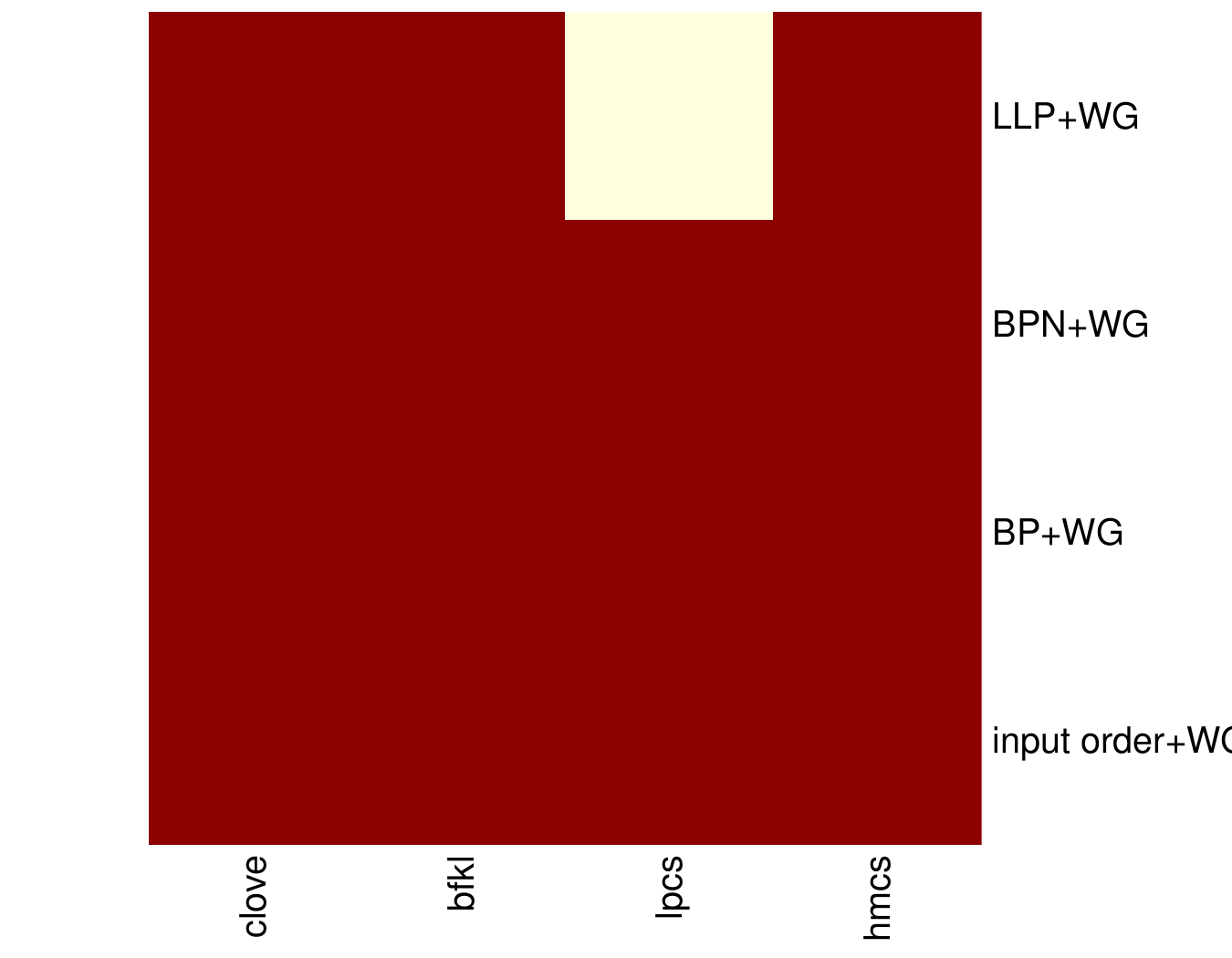}
  \subcaption{Large world}
\end{minipage}

\begin{minipage}[b]{.5\columnwidth} \noindent
  \includegraphics[width=\textwidth]{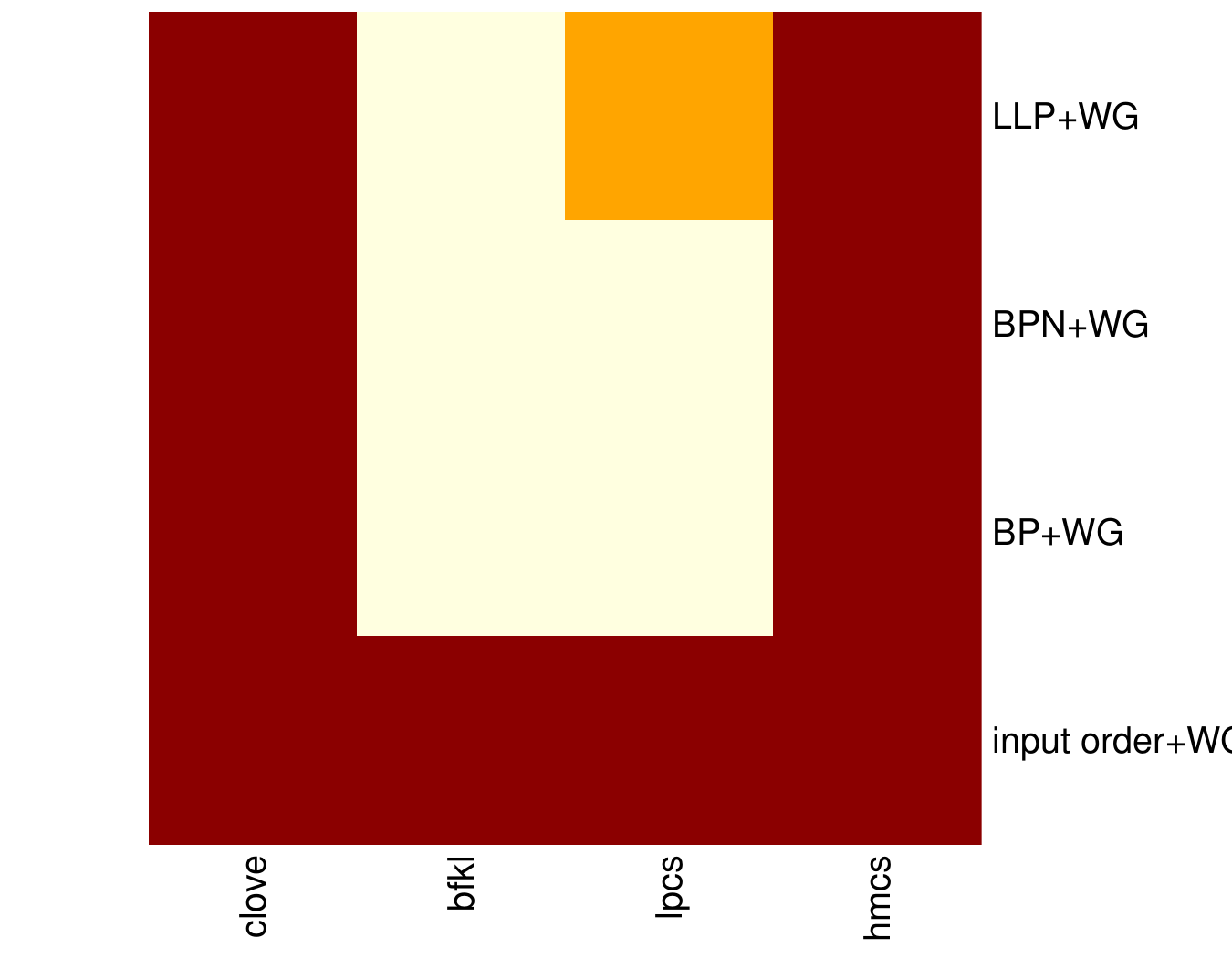}
  \subcaption{Small world}
\end{minipage}
\begin{minipage}[b]{.5\columnwidth} \noindent
  \includegraphics[width=\textwidth]{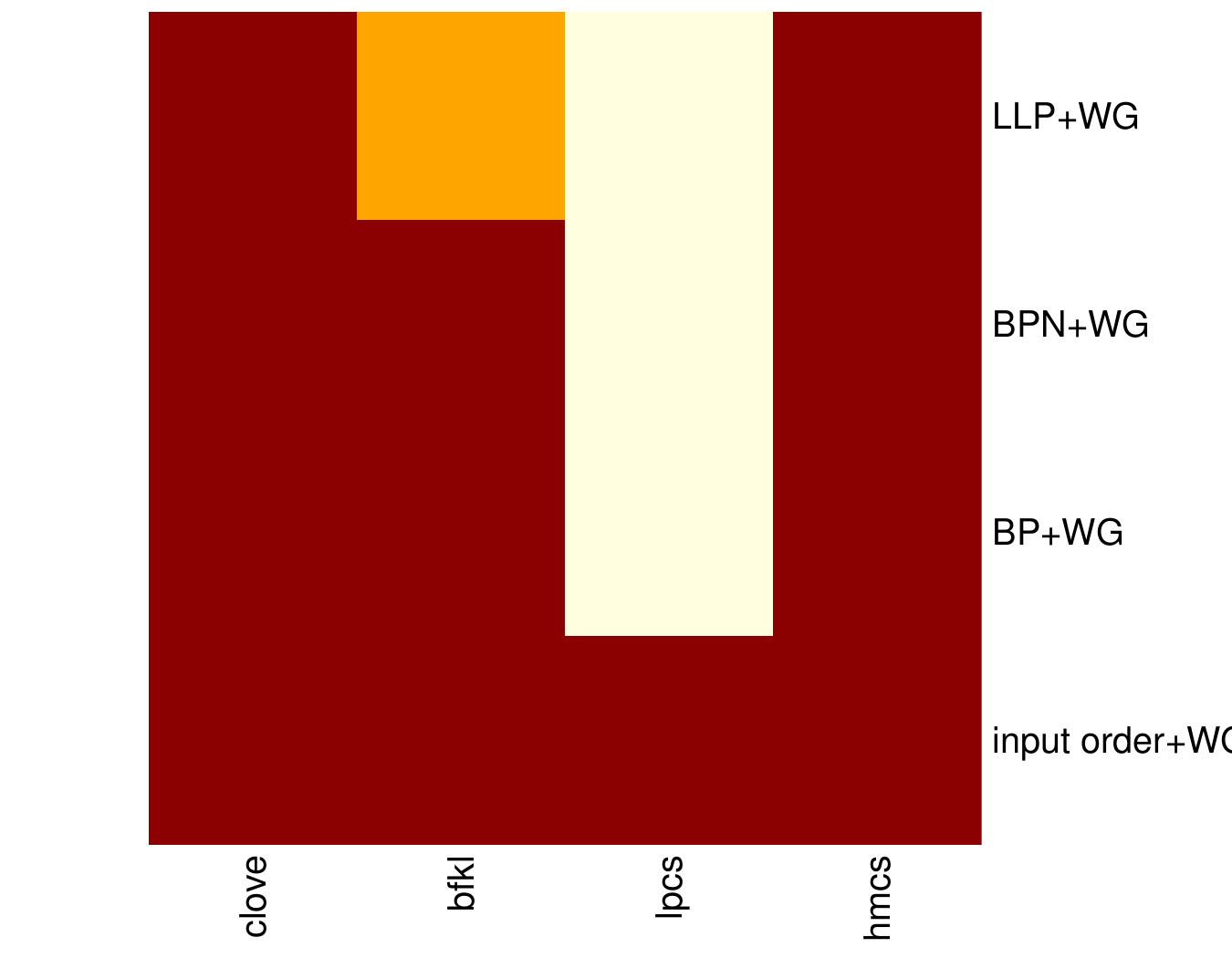}
  \subcaption{Ultra-small world}
\end{minipage}
  \caption{Comparison of the goodness of fit between WebGraph reordering based on hyperbolic embedders and earlier methods. Red suggests that hyperbolic embedders yields better results and the difference is significant; orange suggests lack of significant difference, and yellow suggests significantly worse results for hyperbolic embedders. \label{fig:longlongwg}}
\end{figure*}

Figures \ref{fig:longlongsa} and \ref{fig:longlongwg} show whether our
compression methods give a significant advantage over earlier methods, for various classes of artificial networks.

\begin{table*}
\resizebox{\textwidth}{!}{
\begin{tabular}{|l|c|c|c|c|c|c|c|c|c|c|c|c|c|c|} \hline
graph & \multicolumn{2}{c|}{WebGraph} & \multicolumn{4}{c|}{WebGraph + embedder} & \multicolumn{4}{c|}{HyperFast (8 buckets)} & \multicolumn{4}{c|}{HyperFast (16 buckets)} \\ \cline{3-15} 
      & BP & LLP & BFKL & LPCS & HMCS & CLOVE & BFKL & LPCS & HMCS & CLOVE & BFKL & LPCS & HMCS & CLOVE \\ \hline 
Drosophila2&\cellcolor[HTML]{66C266}$\underset{(0.56,1.16)}{0.86\%}$&\cellcolor[HTML]{3EAA3E}$\underset{(0.24,0.70)}{0.47\%}$&\cellcolor[HTML]{9DDF9D}$\underset{(0.58,1.53)}{1.05\%}$&\cellcolor[HTML]{86D386}$\underset{(0.58,1.46)}{1.02\%}$&\cellcolor[HTML]{108A10}$\underset{(0.14,0.33)}{0.23\%}$&\cellcolor[HTML]{088508}$\underset{(0.12,0.32)}{0.22\%}$&\cellcolor[HTML]{79CC79}$\underset{(0.64,1.32)}{0.98\%}$&\cellcolor[HTML]{FFFFFF}$\underset{(0.97,2.06)}{1.52\%}$&\cellcolor[HTML]{008000}$\underset{(0.14,0.28)}{0.21\%}$&\cellcolor[HTML]{30A030}$\underset{(0.19,0.56)}{0.38\%}$&\cellcolor[HTML]{90D890}$\underset{(0.67,1.38)}{1.02\%}$&\cellcolor[HTML]{DBF7DB}$\underset{(0.89,2.03)}{1.46\%}$&\cellcolor[HTML]{259925}$\underset{(0.17,0.48)}{0.32\%}$&\cellcolor[HTML]{179017}$\underset{(0.11,0.37)}{0.24\%}$\\ \hline 
Human1&\cellcolor[HTML]{1D941D}$\underset{(0.40,0.70)}{0.55\%}$&\cellcolor[HTML]{0D880D}$\underset{(0.22,0.79)}{0.50\%}$&\cellcolor[HTML]{36A436}$\underset{(0.62,1.22)}{0.92\%}$&\cellcolor[HTML]{4BB24B}$\underset{(0.77,1.59)}{1.18\%}$&\cellcolor[HTML]{008000}$\underset{(0.19,0.57)}{0.38\%}$&\cellcolor[HTML]{158E15}$\underset{(0.30,0.77)}{0.54\%}$&\cellcolor[HTML]{77CB77}$\underset{(0.66,2.38)}{1.52\%}$&\cellcolor[HTML]{FFFFFF}$\underset{(1.24,3.50)}{2.37\%}$&\cellcolor[HTML]{67C267}$\underset{(0.52,2.27)}{1.40\%}$&\cellcolor[HTML]{8FD78F}$\underset{(1.09,2.40)}{1.75\%}$&\cellcolor[HTML]{A2E1A2}$\underset{(1.00,2.78)}{1.89\%}$&\cellcolor[HTML]{D7F6D7}$\underset{(1.27,3.22)}{2.24\%}$&\cellcolor[HTML]{56B856}$\underset{(0.51,1.93)}{1.22\%}$&\cellcolor[HTML]{B3E8B3}$\underset{(1.29,2.66)}{1.97\%}$\\
\hline 
\end{tabular}

}
\caption{Coefficients of variation \label{tab:cv}}
\end{table*}



\begin{table}
\resizebox{.9\textwidth}{!}{
\begin{tabular}{|l|c|c|c|c|c|c|c|c|c|c|c|c|c|c|c|c|c|} \hline
graph & trivial & \multicolumn{4}{c|}{WebGraph} & \multicolumn{4}{c|}{WebGraph + embedder} & \multicolumn{4}{c|}{HyperFast (8 buckets)} & \multicolumn{4}{c|}{HyperFast (16 buckets)} \\ \cline{2-18} 
      &         & input & BPN & BP & LLP & BFKL & LPCS & HMCS & CLOVE & BFKL & LPCS & HMCS & CLOVE & BFKL & LPCS & HMCS & CLOVE \\ \hline 
Drosophila2&\cellcolor[HTML]{FFFFFF}$\underset{(8.90,8.90)}{8.90}$&\cellcolor[HTML]{A3E1A3}$\underset{(7.81,7.81)}{7.81}$&\cellcolor[HTML]{89D589}$\underset{(7.52,7.52)}{7.52}$&\cellcolor[HTML]{93DA93}$\underset{(7.54,7.64)}{7.59}$&\cellcolor[HTML]{75CA75}$\underset{(7.32,7.37)}{7.35}$&\cellcolor[HTML]{B0E7B0}$\underset{(7.85,7.97)}{7.91}$&\cellcolor[HTML]{BDECBD}$\underset{(7.94,8.06)}{8.00}$&\cellcolor[HTML]{7DCE7D}$\underset{(7.36,7.38)}{7.37}$&\cellcolor[HTML]{6DC66D}$\underset{(7.32,7.35)}{7.33}$&\cellcolor[HTML]{249924}$\underset{(5.53,5.61)}{5.57}$&\cellcolor[HTML]{38A638}$\underset{(5.80,5.93)}{5.87}$&\cellcolor[HTML]{0C880C}$\underset{(5.03,5.04)}{5.04}$&\cellcolor[HTML]{068406}$\underset{(4.99,5.02)}{5.01}$&\cellcolor[HTML]{2C9D2C}$\underset{(5.57,5.66)}{5.62}$&\cellcolor[HTML]{40AB40}$\underset{(5.87,5.99)}{5.93}$&\cellcolor[HTML]{128C12}$\underset{(5.06,5.08)}{5.07}$&\cellcolor[HTML]{008000}$\underset{(5.00,5.01)}{5.00}$\\
\hline 
\end{tabular}

}
\caption{Compression of real-world graphs: bits per link\label{tab:rw:bpl}}
\end{table}
\begin{table}
\resizebox{.9\textwidth}{!}{
\begin{tabular}{|l|c|c|c|c|c|c|c|c|c|c|c|c|c|c|c|c|} \hline
graph & \multicolumn{4}{c|}{WebGraph} & \multicolumn{4}{c|}{WebGraph + embedder} & \multicolumn{4}{c|}{HyperFast (8 buckets)} & \multicolumn{4}{c|}{HyperFast (16 buckets)} \\ \cline{3-17} 
      & input & BPN & BP & LLP & BFKL & LPCS & HMCS & CLOVE & BFKL & LPCS & HMCS & CLOVE & BFKL & LPCS & HMCS & CLOVE \\ \hline 
Drosophila2&\cellcolor[HTML]{008000}$\underset{(0.14,0.15)}{0.14\mbox{s}}$&\cellcolor[HTML]{0E890E}$\underset{(0.20,0.20)}{0.20\mbox{s}}$&\cellcolor[HTML]{179017}$\underset{(0.22,0.22)}{0.22\mbox{s}}$&\cellcolor[HTML]{40AB40}$\underset{(0.80,0.81)}{0.80\mbox{s}}$&\cellcolor[HTML]{FFFFFF}$\underset{(4.82,5.54)}{5.18\mbox{s}}$&\cellcolor[HTML]{6EC66E}$\underset{(1.14,1.15)}{1.14\mbox{s}}$&\cellcolor[HTML]{7ACD7A}$\underset{(1.14,1.42)}{1.28\mbox{s}}$&\cellcolor[HTML]{B0E7B0}$\underset{(2.34,3.14)}{2.74\mbox{s}}$&\cellcolor[HTML]{E5FAE5}$\underset{(4.62,5.42)}{5.02\mbox{s}}$&\cellcolor[HTML]{55B755}$\underset{(1.02,1.02)}{1.02\mbox{s}}$&\cellcolor[HTML]{4EB34E}$\underset{(1.02,1.02)}{1.02\mbox{s}}$&\cellcolor[HTML]{A5E2A5}$\underset{(2.32,3.02)}{2.67\mbox{s}}$&\cellcolor[HTML]{D7F6D7}$\underset{(4.63,5.33)}{4.98\mbox{s}}$&\cellcolor[HTML]{64C064}$\underset{(1.03,1.03)}{1.03\mbox{s}}$&\cellcolor[HTML]{5CBC5C}$\underset{(1.03,1.03)}{1.03\mbox{s}}$&\cellcolor[HTML]{9BDE9B}$\underset{(2.25,3.03)}{2.64\mbox{s}}$\\
\hline 
\end{tabular}

}
\caption{Compression of real-world graphs: compression time\label{tab:rw:time}}
\end{table}

\section{Implementation details}

Our benchmarking tool is based on the comparison framework from \cite{bridging_iclr}.

We use the C++ reimplementation of LPCS and HMCS from \cite{bridging_iclr}. In the case of HMCS, we use 5 levels of subcommunities.
Since BP \cite{rgb_bp} has no official implementation, we use our own reimplementation in C++. We use the Rust implementation
of WebGraph and LLP \cite{webgraph_rs,webgraph_llp}.

Since we are focused on high compression rather than random access, for WebGraph, we use the hyperparameters used on the
WebGraph website for high compression (window size 16, reference limit 2147483647). Likewise, we do not use
the methods to allow quick random access to HyperFast compressed data, outlined in Subsection \ref{sec:randacc}.

The code-and-data attachment to this paper is available under the anonymized link
\url{https://figshare.com/s/1b04cb24f395b8459790}.

\clearpage

\section{Conclusions and future work}

We have shown that the HyperFast compression method, combined with a fast hyperbolic embedder such as HMCS, yields up to 42\% improvement in the compression
ratio compared to LLP+WebGraph. We conclude with some directions for further research.

\begin{itemize}
\item The CLOVE embedder often yields great compression, but our experiments show it is slower on large graphs than we expected. For compression
applications it might be worthwhile to find out the cause for the slowdown and suggest a compromise solution.

\item HyperFast assumes that the nodes are uniformly distributed along the circle. It is possible that a non-uniform distribution could yield better compression for some graphs.

\item All the methods discussed in this paper use a one-dimensional model of the similarity space. The structure of some graphs is better explained by higher-dimensional
models; there is some research in higher-dimensional hyperbolic embedders \cite{mlembed1,nickel,dmercator,solvbrain_ecai_bez_dziadow}.
However, it appears that such higher-dimensional embeddings require more bits to transfer the embedding itself \cite{bridging_iclr} and the algorithmic techniques used by fast embedders,
WebGraph and HyperFast do not easily carry over to higher dimensions (all the higher-dimensional embedders we know are slow). These issues might limit the applicability in compression.
\end{itemize}


\def\ext{}
\bibliographystyle{plain}
\bibliography{../master.bib}

\end{document}